\documentclass[onecolumn,prb,10pt]{revtex4}

\usepackage{amsmath}   
\usepackage{graphicx}   
\usepackage{caption}
\usepackage{subcaption}  
\usepackage{hyperref}  
\usepackage{amsmath}
\usepackage{amssymb}

\newcommand{\f}{\tau}
\newcommand{\s}{\sigma}
\newcommand{\dd}{\mathrm{d}}
\newcommand{\dk}{D^{(q)}}

\newcommand{\dq}{D^{(0)}}
\newcommand{\dw}{D^{(1)}}
\newcommand{\de}{D^{(2)}}

\newcommand{\ztn}{Z_2^{\otimes n}}
\newcommand{\hh}{\tilde{h}}
\newcommand{\tilt}{\tilde{T}_n^{(\ell)}}
\newcommand{\tnq}{T_{n,q}}

\DeclareMathOperator{\Tr}{Tr}

\DeclareMathOperator{\sym}{SYM}

\DeclareMathOperator\atanh{atanh}

\DeclareMathOperator*{\tr}{Tr}

\begin{document}
\title{One-dimensional disordered Ising models by replica and cavity methods}
\author{C. Lucibello}
\author{F. Morone}
\affiliation{Dipartimento di Fisica, Universit\`a ``La Sapienza'', P.le A. Moro 2, I-00185, Rome, Italy}
\author{T. Rizzo}
\affiliation{CNR-IPCF, UOS Roma Kerberos, Dip. Fisica, Univ. ``La Sapienza'', P.le A. Moro 2, I-00185, Rome, Italy}

\begin{abstract}
Using a formalism based on the spectral decomposition of the replicated transfer matrix for disordered Ising models, we obtain several results that apply both to isolated one-dimensional systems and to locally tree-like graph and factor graph (p-spin) ensembles. We present exact analytical expressions, which can be efficiently approximated numerically, for many types of correlation functions and for the average free energies of open and closed finite chains.
All the results achieved, with the exception of those involving closed chains, are then rigorously derived without replicas, using a probabilistic approach with the same flavour of cavity method.
\end{abstract}

\maketitle

\section{Introduction}

The study of one-dimensional Ising chain with random bonds and/or fields has a long tradition in the context of disordered systems.  Over the years this field has experienced an interesting change in perspective. Earlier studies were essentially motivated by the need to obtain solvable version of three-dimensional models \cite{Rujan78,Derrida78,Derrida83,Grinstein83} and this line of research culminated with the introduction of specific random Hamiltonians that are actually solvable analytically \cite{Nieuwenhuizen86,Luck89,Luck91,Derrida86}.
In the last twenty years dynamical approaches have also been considered as alternatives to equilibrium approaches \cite{Forgacs84,Skantzos00,Fisher01,Coolen12} while developments in the context of static studies \cite{Monasson96,Nikoletopoulos2004}  have been mainly motivated by the connection between one dimensional systems and models defined on sparse random graphs. Random graphs in turn have many important applications in the context of computer science, artificial intelligence and information theory \cite{Parisi1987,Montanari2009}. In this broader context one is more interested in having a general formalism that can be applied to any given distribution of the quenched Hamiltonians at the price of obtaining the result through numerical solution of implicit equations.

In the general case one would like to study an Ising chain, either open or closed, of arbitrary length $L$ were the fields and couplings are i.i.d. random variables. Quantities of interest include the free energy but also all sort of averaged correlation functions. Indeed, at variance with pure systems, correlations can be averaged in two different ways: over thermal noise (conventionally referred as {\it connected correlations}) and over the quenched Hamiltonians ({\it disconnected correlations}). This difference is important both at the theoretical and the practical level. Indeed disconnected correlations happen to be  much larger in random field systems (but not in Spin-Glasses) and lead to a very complex phenomenology, {\it e.g.} the increase in the critical dimension from $D=4$ of the pure ferromagnet to $D=6$ \cite{Young98}.   
In this paper we show how to complete this program  by means of the replica method, more precisely by means of the replicated transfer matrix (RTM) approach.
As long as the sources of disorder are independently distributed, one can express the integer moments of the partition function through traces of powers  of the $2^n\times 2^n$ transfer matrix of a system of $n$ replicated spins. Then, as usual with replica calculations, the analytic continuation to $n=0$ is performed.
We will derive expressions for the aforementioned quantities in terms of the solutions of integral equations that can be solved for instance through population dynamics algorithms. 
In order to do so we build on the crucial contribution of Monasson and Weigt \cite{Monasson96}, who first characterized the spectral properties of the RTM. 
The motivation is not only to have a compilation of useful formulas but also to present  some non-trivial features of their derivation. The most important is connected with the fact that in the limit $n \rightarrow 0$ two families of eigenvalues (corresponding to the Longitudinal and Anomalous  sectors in the Spin-Glass jargon) become degenerate. From the theoretical perspective, this determines an anomalous behaviour of the disconnected correlation functions and of corrections to the free energy of closed chains. On a practical side this implies that one has to determine not only the eigenvalues and eigenvectors of the integral equations at $n=0$ limit but also their first $O(n)$ correction.

While the replica method is at present the only way to derive expressions for  all quantities of interest in a compact form,  its well-known drawback is the assumption that one can make the  
analytical continuation $n \rightarrow 0$ of expressions whose derivation makes sense only for positive integer $n$. One is therefore interested in deriving the same expressions in a more direct way.
Unfortunately there are no general results or strategies on how to do this and one has to proceed case by case.
We will present a direct probabilistic derivation of many of the expressions obtained through the replica method. A particularly non-trivial result is the derivation of the formula for disconnected correlation functions that has been long sought for. Such a derivation is based on the fact that a direct physical meaning can be attributed  to the continuation of replica expressions to real $n$, at variance with other classic analytical continuation tricks ({\it e.g.} dimensional continuation in field theory). Therefore one can first derive rigorously an expression at any real $n$ and then safely take the limit $n \rightarrow 0$.
The only replica expression whose derivation is left as an open problem is the free energy of closed chains. We recall that closed chains are rather important objects that appears in perturbative computations developed around the tree approximation \cite{Ferrari2013}. 

We conclude this introduction by briefly  discussing the connection between our results and the extensive literature on disordered Ising chains. As we said already earlier studies, appeared in the context of the random field Ising model (RFIM), were motivated essentially by the possibility of obtaining exact solutions when dealing with one-dimensional models. It was immediately recognized  \cite{Rujan78,Derrida78,Derrida83} that the free energy of an infinite chain can be expressed in terms of an iterative equation which corresponds to the Longitudinal sector in the terminology of RTM. Exact results can be obtained at zero temperature, where the equation can be solved explicitly \cite{Derrida78} or by enumeration in the special case in which the random fields are either zero or infinity \cite{Grinstein83}. Finally it was discovered that some models with specific distributions of the disorder can be solved analytically \cite{Nieuwenhuizen86,Luck89,Luck91,Derrida86}. 
Much effort has been put in the study of the iterative equation relevant to the free energy of the infinite chain, starting from the observation that when the random fields and couplings take a discrete number of values ({\it e.g.} $H=\pm 1$) the solutions of the equations may display a multi-fractal structure \cite{Gyorgyi84,Aeppli83,Bruinsma1983,Behn88}. 
The approaches taken  to characterize the correlation functions have been less successful. Connected correlations where computed exactly for the aforementioned solvable models but the disconnected correlations resisted all efforts \cite{Luck89} to capture their expected peculiar features (the double pole) \cite{Young98} up to this work. Later on correlations functions were also studied in a more general framework at zero temperature \cite{Igloi94}, but again non considered the disconnected correlations. The results of \cite{Luck89} are system-specific and not based on iterative methods, only recently \cite{Janzen2010,Janzen2010a,Parisi14} it has been recognized that general iterative expressions for connected correlations can be obtained by means of cavity arguments like those we will present in the following. 

The paper is organized as follows: in Section \ref{sec:main} we define the model we are considering and expose the main results of this paper; in Section \ref{sec-decomp} we develop all the spectral formalism of the RTM and we apply it in Section \ref{sec:applications} to the computation of free energies and correlation functions. Most of the results obtained with the RTM are then rederived with a purely probabilistic approach in Section \ref{sec:cavity}.

\section{Definitions and main results}
\label{sec:main}

\begin{figure}
\centering
\includegraphics[width=\textwidth]{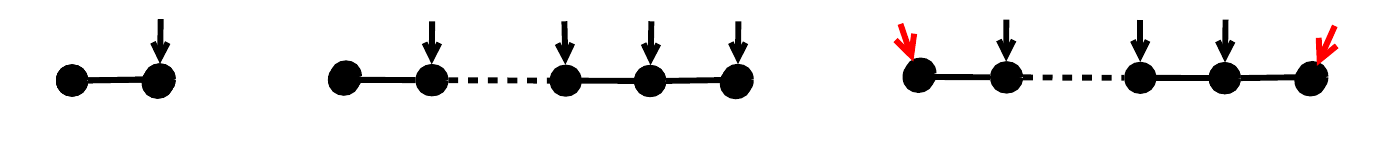}
\caption{Pictorial representation of the matrix $T_n$ (\textit{left}), its powers $T_n^\ell$ (\textit{center}) and the matrix $\tilt$ (\textit{right}).}
\label{fig:ttt}
\end{figure}
In this paper we consider one-dimensional Ising spin system with i.i.d. random fields  and couplings , e.g. an isolated chain or a chain embedded in a locally tree-like graph, therefore  described by the product of uncorrelated $2\times 2$ random transfer matrices $M_i$ defined by 
\begin{equation}
M_i(\s_{i+1},\s_{i}) = e^{\beta J_i \s_{i+1}\s_{i}  + h_i \s_i}.
\end{equation}
The partition function of a closed chain of length $\ell$ is then a random variable given by 
\begin{equation}
Z_{\ell,c}= \Tr\,\prod_{i=0}^{\ell-1} M_i .
\end{equation}
A powerful technique to compute the statistical properties of this kind of objects is the well known replica method \cite{Parisi1987}. As we shall see, as long as the system stays in a replica symmetric phase, its statistical properties are encoded in the (replica symmetric) replicated transfer matrix $T_n$, the 
$2^n\times 2^n$ matrix defined by
\begin{equation}
T_n(\s, \f)=\mathbb{E}_{J,h}\ \mathrm{e}^{\beta J\sum_{a=1}^n\s^a \f^a+\beta h\sum_{a=1}^n\f^a} .
\label{TM}
\end{equation}
Here and in the following we denote with $\s$  the vector $(\s^1,\ldots,\s^n)$,  with the $n$ replicated spins $\s^a$ taking values in $Z_2=\{-1,1\}$. A similar definition holds for $\f$. As usual in the replica method\cite{Parisi1987} we shall work at integer value of $n$ and perform the analytic continuation for $n\downarrow 0$ at the end of the computations. We shall assume in the following that the field $h$ is an arbitrary distributed external random field, if we are considering an isolated chain, or, if we are considering a chain embedded in a locally tree-like graph, to be a random cavity field conditioned to act on a spin that is already connected to two other spins (its neighbours on the chain). See Figure \ref{fig:ttt} for a representation of $T_n$ and its powers $T_n^\ell$.

A first spectral analysis of $T_n$ was conducted by Weigt and Monasson \cite{Monasson96}. Following their lead we take advantage of the replica index permutation symmetry of $T_n$ to choose an appropriate bases to express its right eigenvectors. There are $n+1$ non-equivalent irreducible representation of the permutation group, which can be glued together to form the \textit{sectors} $\dk,\ q=0,1,\ldots,\left\lfloor \frac{n}{2}\right\rfloor,$ partitioning $\ztn$. In the following, with some abuse of notation, we will denote with $\dk$ the set of eigenvalues of $T_n$ with eigenvector in that sector.  The eigenvectors of $T_n$ in the sector $\dk$ can be parametrized by functions $g^\lambda_q(u)$ that, in the limit $n\downarrow 0$, satisfy the eigenvalue equation
\begin{equation}
\lambda\, g^\lambda_q(u)=\mathbb{E}_{J,h}\int \dd v\ g^\lambda_q(v)\ \delta\left(u-\hat{u}(J,v+h)\right)\ \left(\frac{1}{\beta}\frac{\partial\hat{u}}{\partial{ v}}\right)^q\ ,
\label{eig-g}
\end{equation}
where $\hat{u}(J,h)=\frac{1}{\beta}\atanh\left(\tanh(\beta J)\tanh(\beta h)\right)$ is the cavity iteration rule.

In this paper we extend the analysis of the spectral properties of $T_n$ to achieve a complete description of the $n\downarrow 0$ limit, derive exact expressions for correlation functions and free energies of chains. Since $T_n$ is the product of two non-singular symmetric matrices, it possess a complete orthonormal (in the left-right sense) basis of left and right eigenvectors with real eigenvalue. The left eigenvector corresponding to a certain right $\psi_R$ is simply $\mathbb{E}_{h}e^{\beta h\sum_a \s^a}\psi_R(\s)\equiv \rho_h(\s)\psi_R(\s)$.
Therefore the spectral decomposition of $T_n$ into the subspaces $\dk$ is given by
\begin{equation}
T_n(\s,\f)=\sum_{q=0}^{\left\lfloor\frac{n}{2}\right\rfloor}\sum_{\lambda\in \dk}\lambda\ \rho_q^{\lambda}(\s)\rho_h(\f)\rho_q^{\lambda}(\f)\sum_{\substack{a_1<\cdots<a_q \\ b_1<\cdots<b_q}}Q_{a_1\dots a_q;b_1\dots b_q}\,\s^{a_1}\dots \s^{a_q}\f^{b_1}\dots \f^{b_q}\ .
\label{tq}
\end{equation}
Here we have denoted with $\rho_q^{\lambda}(\f)$ the replica symmetric part of the eigenvector in the sector$\dk$ with eigenvalue $\lambda$. The second sum is over all the eigenvalues of $T_n$ in the sector $\dk$, given in the $n\downarrow 0$ limit by the solutions of Eq. \eqref{eig-g}. The coefficients $Q_{a_1\dots a_q;b_1\dots b_q}$ have simple algebraic expressions in each sector (see Eqs. \eqref{Q1} and \eqref{Q2})  and are invariant under  permutations of any of their two sets of indices.
While a different left-right decomposition of $T_n$ has already been attempted\cite{Nikoletopoulos2004}, an unfortunate choice in the parametrization of the eigenvectors in terms of function of two variables led to an unmanageable formalism.  Thanks to the spectral representation \eqref{tq} we can easily take the powers of $T_n$ and contract the matrix with the quantities we want to average. In Section \ref{sec-decomp} we derive Eqs. \eqref{eig-g} and \eqref{tq}, and discuss the non-trivial aspects of the small $n$ limit.

One of the applications of the spectral formalism is the computation of the average free energy of open and closed chains, as exposed in Section \ref{subsec:chains}. Recently it has been shown \cite{Ferrari2013} that the first finite size correction to thermodynamic free energy of systems on diluted graphs can be expressed as a linear combination of the free energies of closed and open chains. It has also been argued \cite{Parisi2012,Ferrari2013} that a perturbative expansion around the Bethe approximation towards finite dimensional lattices, shall account for the presence of loops (closed chains) and will contain the free energies and the correlation function of one-dimensional objects, motivating the importance of exact and easily approximable expressions for their free energies.

 Taking the trace of $T_n^\ell$ and performing the $n\downarrow 0$ limit one obtains the average free energy of a closed chain of size $\ell$:
\begin{equation}
-\beta f_{\ell}^{c}=-\beta\ell  f_0
+\sum_{\lambda\in\dw}\Delta_\lambda\,\ell\,\lambda^{\ell-1}
+\sum_{q=1}^{\infty}\hat{d}_q
\sum_{\lambda\in\dk} \lambda^{\ell}\ .
\label{fclo}
\end{equation}
The non-trivial features of this expression is the presence of a term $O(\ell\,\lambda^{\ell-1})$. This is typically not present in a ordinary eigenvalue decomposition that contains only $O(\lambda^{\ell})$. Its presence is due to the $n \rightarrow 0$ limit combined with the fact that the longitudinal and anomalous eigenvalues become degenerate. As we said in the introduction this is a phenomenon that has dramatic  physical consequences in the RFIM context \cite{Young98}.

The terms  $\Delta_\lambda$, due to the degeneracy between the eigenvalues of the sectors $\dq$ and $\dw$ at $n=0$, are expressed in Eq. \eqref{Delta}. The coefficients $\hat{d}_q$ are the analytic continuation of the degeneracies of the eigenvalues, and are given in Eq. \eqref{dhat}. We note that the correction to the intensive free energy $f_0$ (expressed in Eq. \eqref{f0}) is given by a linear combination of exponential and $\ell$ times exponential terms. The decaying part of   $f_{\ell}^{c}$ is dominated by the largest eigenvalue among the various sectors.

In the computation of the free energy of an open chain, we allow for the incoming fields at the extremities of the chain to be distributed differently from the fields $h$ acting on the internal spins, and denote them by $\hh$. This is in fact what happens in general when considering an open chain embedded in a sparse graph. The expression we derived for the average free energy of an open chain of length $\ell$ is 
\begin{equation}
\begin{aligned}
-\beta f^o_{\ell}=& -\ell\beta f_0+\mathbb{E}_{\hh}\int \dd u\ P(u)\,2\log\cosh\left(\beta( u+\hh)\right)-\mathbb{E}_{h}\int \dd u\dd v\ P(u)P(v)\log\cosh\left(\beta( u+v+h)\right)\\
&+\log 2 +\sum_{\lambda\in D^{(1)}} a_{\lambda,0}^2\,\lambda^\ell \ , 
\end{aligned}
\label{fop}
\end{equation}
where $P(u)$ is the distribution of cavity messages along the chain and the coefficients $a_{\lambda,0}$ are related to the left eigenvectors of the sector $\dq$ and given in Eq. \eqref{b-lambda}.

Another result we will present is the expression of the connected correlation functions of two spin at distance $\ell$, in a form that is both analytically exact and easy to approximate numerically with high precision. In Section \ref{subsec-corr} we derive the formula 
\begin{equation}
\overline{\langle \s_0\s_{\ell}\rangle_{\mathrm{c}}^q}=\ \sum_{\lambda\in\dk}a^2_{\lambda,q}\, \lambda^{\ell} \ ,
\label{cq}
\end{equation}
where $a_{\lambda,k}$ can be computed through the eigenfunction $g_q^\lambda$ using Eq. \eqref{alk}. We indicate with $\overline\bullet$ the average over all kinds of disorder in the model considered. For Ising model on sparse random graphs with mean residual degree $z$, the susceptibility  $\chi_q =\sum_{i<j} \frac{1}{N}\mathbb{E}\langle \s_i\s_j\rangle_{\mathrm{c}}^q $ diverges when the greatest eigenvalue of $\dk$ reaches the value $1/z$. Therefore the sectors $\dw$ and $\de$ are the relevant ones to the ferromagnetic and the spin-glass transitions respectively (see Figures \ref{img:eig1} and \ref{img:eig2}).

The computation of the thermally disconnected correlation function, 
$\overline{\langle\s_0\rangle\langle \s_\ell\rangle}- \overline{\langle\s_0 \rangle}\,\overline{\langle\s_\ell \rangle}$, particularly relevant to the RFIM \cite{DeDominicis2006}, requires a careful treatment of the analytic continuation to $n=0$. The final expression we obtained, Eq. \eqref{c-disc3}, is not a linear combination of terms involving only one eigenvalue, as in the previous formulas. The leading term for large $\ell$ is easily extracted though: let $\lambda_1$ be the greatest eigenvalue of the sector $\dw$, then
\begin{equation}
\overline{\langle\s_0\rangle\langle \s_\ell\rangle}- \overline{\langle\s_0 \rangle}\,\overline{\langle\s_\ell \rangle} =
\Delta_{\lambda_1}\, a^2_{\lambda_1,1}\,\ell\,\lambda_1^{\ell-1}
+O(\lambda_1^\ell) \qquad \text{for } \ell\to+\infty,
\label{c-disc}
\end{equation}
with $\Delta_\lambda$ and $a_{\lambda,1}$ given in Eq. \eqref{Delta} and Eq. \eqref{alk} respectively.  Therefore, on one-dimensional chains and sparse graphs,  the susceptibility corresponding to the thermally disconnected correlation function  present the characteristic double pole behaviour near the transition point, whose prefactor can also be computed by Eq. \eqref{c-disc}. 

The expressions we found for free energies of chains, Eqs. \eqref{fclo} and \eqref{fop}, and the correlation function Eqs. \eqref{cq} and \eqref{c-disc}, are exact for every value of the length $\ell$ of the chain but involve the computation of infinitely many terms. Fortunately it turns out from our numerical simulations that the spectrum of the integral operator in Eq. \eqref{eig-g} is discrete and the eigenvalues are well spaced. Therefore considering only the first few highest eigenvalues one obtains very good approximations already at small values of $\ell$. They can be computed numerically, discretizing the kernel of the integral operator of Eq. \eqref{eig-g} and directly computing the eigenvalues of the associated matrix. Moreover the leading eigenvector and eigenvalue of each sector can be efficiently selected with multiple applications of the discretized operator on an arbitrarily chosen vector (as it was done to obtain Figures \ref{img:eig1} and \ref{img:eig2}).

All the results we obtained using the replicated partition function formalism, with the noticeable exception of the formula for the average free energies of closed chains Eq. \eqref{fclo}, can be recovered using a purely probabilistic approach in the same spirit of the usual cavity method \cite{Parisi1987}\cite{Montanari2009}.
 
In Section \ref{sub-cav-open} we devise two alternative probabilistic derivation for the average free energies of  open chains . The first is based on a recursive equation involving the moments of the partition function, which leads to an expression for the moment of the random partition function $Z^n_{\ell}$ of an asymmetric open chain in terms of the left and right eigenvector of an integral operator we also encountered in the RTM formalism: 
\begin{equation}
\overline{Z^n_{\ell}(u;x)}=\sum_{\lambda\in\dq}\lambda^\ell(n)\, g_0^\lambda(u;n)\, S_0^\lambda(x;n)\, [2\cosh(\beta x)]^n.
\end{equation}
Here $n$ is not related to the number of replicas, since replicas are not present in this approach, but is an arbitrarily chosen real positive number.
The other method presented in Section \ref{sub-cav-open} the iteration of the average free energy itself during the construction of the chain, which requires to keep track of the message of $u_\ell$ the cavity message propagating through a chain at distance $\ell$ from one of the extremities, at each iteration. The two approaches are deeply related and obviously lead to the same result.

Crucial to the probabilistic computation of the connected correlation functions, ah has been noted recently\cite{Janzen2010a} is the random variable $X_\ell$ defined by $X_\ell\equiv \frac{\partial u_\ell}{\partial H_0}$, where$ H_0$ field acting on the same extremity. It turns out that the connected correlation function of Eq. \eqref{cq} is encoded in the $q$-the moments of the joint law of $u_\ell$ and $X_\ell$ at fixed $u_\ell$. This object, the function 
\begin{equation}
G_q^{(\ell)}(u)=\int \dd X\ P_{\ell}(u,X)\ X^q,
\end{equation}
obeys a recursion rule, Eq. \eqref{rec-gq}, containing the integral operator of Eq. \eqref{eig-g}.  Expressing $G_q^{(\ell)}(u)$ in the basis of the eigenvalues of $\dk$ leads then straightforwardly to the expression \eqref{cq} we obtained using replicas. Moreover in Section \ref{subsec:cav-conn} a more general result is presented in Eq. \eqref{conn-cav-zn}. 

The thermally disconnected correlation function  $\overline{\s_0\s_{\ell}\rangle_c}$ is computed in Section \ref{sub-cav-disc}, using some results we obtained for the connected correlation function and for the moments of the partition function of an open chain, thanks to the relation
\begin{equation}
\frac{\partial}{\partial H_0}\frac{\partial}{\partial H_\ell} Z^n_{\ell,o} = 
n\, \langle \s_0\s_{\ell}\rangle_c \ Z^n_{\ell,o} + n^2\, \langle \s_0\rangle\langle\s_{\ell}\rangle\ Z^n_{\ell,o}.
\end{equation}

An alternative approach, technically more difficult, outlined in Section \ref{sub-cav-disc} involves the resolution of an iterative equation for the function $R^{(\ell)}(u)=\overline{\delta(u-u_\ell)\langle\s_0\rangle^{(\ell)}}$, which takes into account the shift in the magnetization of the  (initial) spin at the other side of the chain with respect to the spin where a new spin is attached to increase the length of the chain.

In the following Sections we fill-in all the technical details associated to the previous claims.

\section{Spectral decomposition}
\label{sec-decomp}
We present an in-depth treatment of the spectral theory of the replica symmetric RTM. In Section \ref{subsec:permut} we discuss the spectral decomposition of the matrix for integer values of the number of replicas $n$. We introduce in Section \ref{subsec:integr} an integral representations of the eigenvectors, in order to discuss the main features of the analytic continuation to small values of $n$ in Section \ref{subsec:smalln}. In Section \ref{subsec:degeneracy} we discuss some technicalities related to a peculiar aspect of the $n\downarrow 0$ limit, the degeneracy between the Longitudinal and the Anomalous sectors. 
\subsection{The Permutation Group}
\label{subsec:permut}
The $2^n\times 2^n$ matrix $T_n$ defined by Eq. \eqref{TM} is invariant under the action of the group of permutations among the replicated spins: for each permutation $\pi$ acting on the $n$ spin, we have the equivalence  $T_n(\pi(\s),\pi(\f))=T_n(\s,\f)$. 
This symmetry  allows us to  block-diagonalize $T_n$ according to the irreducible representations of the permutation group. This idea has been first introduced by Weigt and Monasson\cite{Monasson96} in order to compute the eigenvalue spectrum of $T_n$.

For the sake of completeness we now review Weigt and Monasson's method,  then we extend it further, in order to achieve the decomposition of the transfer matrix in terms of left and right eigenvectors. 
The replicated space is $\ztn$. Let's call $\Delta_m$, with $m=0,...,n$, the subspace of configurations having exactly $m$ spins up. These subspaces are clearly invariant under any permutation of the replicas, therefore we can consider the representation of the permutation group in the $n+1$ subspaces $\Delta_m$ and look for the irreducible ones. The complete decomposition of $\Delta_m$ into irreducible subspaces $D^{(m,q)}$ has been done by Wigner\cite{Wigner1959}. It reads:
\begin{equation}
\begin{aligned}
&\Delta_0 = D^{(0,0)} \ , \\
&\Delta_1 = D^{(1,0)} \oplus D^{(1,1)}\ ,\\
&\ldots \\
&\Delta_m = D^{(m,0)} \oplus\ldots\oplus D^{(m,\ \min(m, n-m))}\ ,\\
&\ldots \\
&\Delta_{n-1} = D^{(n-1,0)} \oplus D^{(n-1,1)}\ ,\\
&\Delta_n = D^{(n,0)}\ . \\ 
\label{dqm}
\end{aligned}
\end{equation}
Representations $D^{(m,q)}$, at fixed $q$, are isomorphic and have dimension
\begin{equation}
d_q\equiv\dim\left(D^{(m,q)}\right)= {n \choose q} - {n \choose q-1}  \qquad q=0,...,\lfloor n/2\rfloor \ ,
\label{dq}
\end{equation}
where $\lfloor x\rfloor$ is the smallest integer part of $x$. Notice that  by definition $d_0=0$.
As we have $(n+1-2q)$ subspaces $D^{(m,q)}$, the $q$-sector of our matrix $T_n$ will contain $(n+1-2q)$ eigenvalues with degeneracy $d_q$. One can check that $\sum_{q=0}^{\lfloor n/2\rfloor} d_q\ (n+1-2q) = 2^n$.

A vector of the space $D^{(m,q)}$ can be constructed using Young tableaux\cite{Janzen2010}, and has the form
\begin{equation}
|m, q \rangle=\left(|+\rangle|-\rangle-|-\rangle|+\rangle\right)^q \sym\left(|+\rangle^{m-q}|-\rangle^{n-m-q}\right)\ , 
\label{eq:basisvect}
\end{equation}
where the operation $\mathrm{SYM}$ means a complete symmetrization with respect to the $n-2q$ last entries (the first $2q$ entries are, instead, anti-symmetrized).
A basis of the subspace $D^{(m,q)}$ can be constructed by applying all the transformations of the permutation group to the vector $|m, q \rangle$ in Eq. \eqref{eq:basisvect} and choosing a maximal linearly independent subset.

We look for the eigenvectors of $T_n$ in the subspaces 
\begin{equation}
\dk = \bigoplus_{m=q}^{n-q} D^{(m,q)}  \qquad q=0,\ldots, \left\lfloor \frac{n}{2}\right\rfloor
\label{defdq}
\end{equation}
of dimension $d_q (n+1-2 q)$. Since $T_n$ has no symmetries beside the replicas permutation one, it has $n+1-2 q$ different eigenvalues in $\dk$, each with multiplicity $d_q$. In the following we will refer to the subspaces $\dk$ as to \textit{sectors}. Moreover, with some abuse of notation, we shall use the symbol $\dk$ for the set of eigenvalues corresponding to eigenvectors in that sector. 

Of particular relevance are the sectors $\dq$, $\dw$ and $D_0^{(2)}$ since they are associated to the Longitudinal, Anomalous and Replicon modes respectively from mean-field spin-glass theory\cite{Bray1979}, as we will later show when discussing correlation functions in Section \ref{subsec-corr}.

By Eqs. \eqref{eq:basisvect} and \eqref{defdq} it is possible to factorize the replica symmetric part in the eigenvectors $\psi^\lambda_q(\s)$ of the transfer matrix in the sector $\dk$, that is we can write
\begin{equation}
\psi^\lambda_q(\s)=\rho^\lambda_q\left(\sum_a\s^a\right)\sum_{a_1<\cdots<a_q}C_{a_1\dots a_q}\,\s^{a_1}\dots \s^{a_q}\ ,
\label{psilq}
\end{equation}
where the replica symmetric part $\rho^\lambda_q$ of the eigenvectors is the one relevant to the computation of the eigenvalues. By last equation the eigenvectors of the sector $\dq$ are completely replica symmetric. The coefficients  $C_{a_1\dots a_q}$ are invariant for any permutation of the indices and are equal to zero if any two of the indices are equal. Moreover they have to satisfy the constraint 
\begin{equation}
\sum_{a_1=1}^n\ C_{a_1\dots a_q}=0,
\label{constr-c}
\end{equation}
which is a necessary and sufficient condition for any vector of the form of Eq. \eqref{psilq} to belong to the subspace $\dk$. Any set of $d_q$ linearly independent coefficient vectors $C$ can be chosen as an appropriate basis for the subspace. It is easy to prove that the product of two non-singular symmetric matrices possess a complete orthonormal (in the left-right sense) basis of left and right eigenvectors with real eigenvalues, and this is indeed case for $T_n$. In fact if we define, with a little abuse of notation, the vector
\begin{equation}
\rho_h(\s)\equiv \mathbb{E}_h e^{\beta h\sum_a  \s^a}\ ,
\end{equation}
than $T_n(\s,\f)=
\sum_{\s'} \mathbb{E}_J\, e^{\beta J \s\s' }\times\delta_{\s'\f}\rho_h(\f)$. Moreover the left eigenvector $\psi_L$ corresponding to a certain right $\psi_R(\sigma ; \lambda,k)$ is  simply given by
\begin{equation}
\psi_L(\s;\lambda,k)=\rho_h(\s)\psi_R(\s;\lambda,k)\ ,
\label{psi-l}
\end{equation}
where $k$ denotes one choice of the coefficients $C_{a_1\dots a_q}$ among the $d_q$ possible. Imposing the orthonormality condition
\begin{equation}
\sum_\sigma \psi_L(\s;\lambda,k)\,\psi_R(\s;\lambda',k')=\delta_{\lambda\lambda'}\,\delta_{k k'}\ ,
\end{equation}
with the sum ranging over all the $2^n$ configuration of the replicated spin, and after successive application of Eq. \eqref{constr-c}, we  obtain
\begin{equation}
\sum_\s\rho^\lambda_q(\s)\rho_h(\s)\rho^{\lambda'}_q(\s)
\prod_{a=1}^q(1-\s^{2a-1}\s^{2a})=\delta_{\lambda\lambda'}
\label{normrho}
\end{equation}
along with
\begin{equation}
\sum_{a_1<\cdots<a_q}C^k_{a_1\dots a_q}C^{k'}_{a_1\dots a_q}=\delta_{k k'}.
\label{normc}
\end{equation}
We are now able to write down the transfer matrix in the spectral form
\begin{equation}
T_n(\s,\f)=\sum_{q=0}^{\left\lfloor\frac{n}{2}\right\rfloor} \tnq(\s,\f)
\label{spectr-dec}
\end{equation}
where $\tnq$ is the restriction of $T_n$ to the subspace $\dk$, defined by
\begin{equation}
\tnq(\s,\f)=
\sum_{\lambda\in\dk}{\lambda}\ \rho_q^{\lambda}(\s)\rho_h(\f)\rho_q^{\lambda}(\f)
\sum_{
\substack{
a_1<\cdots<a_q  \\
b_1<\cdots<b_q 
}}
Q_{a_1\dots a_q;b_1\dots b_q}\,
\s^{a_1}\dots\s^{a_q}\f^{b_1}\dots\f^{b_q}\ .
\end{equation}
The coefficients $Q$ appearing in last expression are invariant for any permutation of the set of indices $a$ or $b$, therefore they depended only on the number of equal indexes in the sets $\{a_1,\ldots,a_q\}$ and $\{b_1,\ldots,b_q\}$. They are defined by
\begin{equation}
Q_{a_1\dots a_q;b_1\dots b_q}=\sum_{k=1}^{d_q}C_{a_1\dots a_q}^kC_{b_1\dots b_q}^k\ ,
\end{equation}
and their $(q+1)$ different values can be computed applying recursively Eqs. \ref{constr-c} and \ref{normc}. If we denote $Q^{(q)}_p$ the coefficient in the sector $\dk$ with $p$ pairs of different indexes, for the first sectors we have 
\begin{align}
Q^{(1)}_0 &= \frac{n-1}{n}  & Q^{(1)}_1 &= -\frac{1}{n}   &  \label{Q1}\\
Q^{(2)}_0 &= \frac{n-3}{2(n-1)}  & Q^{(2)}_1 &= -\frac{Q^{(2)}_0}{n-2}   &Q^{(2)}_2 = -\frac{2 Q^{(2)}_1}{n-3}
\label{Q2}
\end{align}

\subsection{Integral representations}
\label{subsec:integr}
In order to perform the limit $n\downarrow 0$ it is convenient to find a suitable parametrization for the eigenvectors of the form \eqref{psilq}. For the replica symmetric part of the eigenvectors $\psi^\lambda_q$, see Eq. \eqref{psilq}, we employ the standard parametrization
\begin{equation}
\rho^\lambda_q(\s)=\int \dd u \ g^\lambda_q(u;n)\, \frac{e^{\beta u \sum_a \sigma^a}}
{\left[2 \cosh(\beta u)\right]^n},
\label{rapr-rho}
\end{equation}
in terms of the functions $g^\lambda_q(u;n)$. Turns out that all the functions g$^\lambda_q$ parameterizing the eigenvectors of the sector $\dq$, are by themselves the eigenfunctions of an integral operator associated to that sector.  In fact, 
expressing the linear terms in Eq. \eqref{psilq} through the identity
\begin{equation}
\s_{a_1}\ldots\s_{a_q}=\left.\frac{\partial}{\partial \epsilon_{a_1}}\ldots\frac{\partial}{\partial \epsilon_{a_q}}\right|_{\epsilon=0}\,e^{\sum_a \epsilon_a \s^a}
\end{equation}
and plugging Eq. \eqref{rapr-rho} into the eigenvalue equation $T_n \psi_q=\lambda\, \psi_q$, we obtain, after some manipulations, the new eigenvalue equation
\begin{equation}
\begin{aligned}
\lambda\ g^\lambda_q(u;n)=\mathbb{E}_{J,h}\int \dd v\ \delta\big(u-\hat{u}(J,h+v)\big)\left(\frac{1}{\beta}\frac{\partial\hat{u}}{\partial{v}}\right)^q Z^n(J,h,v)\  g^\lambda_q(v;n).
\label{eqeig-n}
\end{aligned}
\end{equation}
The function $\hat{u}(J,x)$, defined by
\begin{equation}
\hat{u}(J,x)=\frac{1}{\beta}\atanh\left(\tanh(\beta J)\tanh(\beta x)\right),
\label{hatu}
\end{equation}
will be recognized by the learned reader as the update rule for cavity messages.
As we shall see, the function 
\begin{equation}
Z(J,h,v)=\frac{2 \cosh(\beta J)\cosh\left( \beta(v+ h)\right)}{\cosh( \beta v)} 
\end{equation}
is related to the intensive free energy of an chain. Notice that in writing down Eq. \eqref{eqeig-n} we have shifted the problem of finding a complete bases of eigenvectors for the matrix $T_n$ to the equivalent problem of the spectral decomposition of the integral operators of Eq. \eqref{eqeig-n}, for $q=0,\ldots,\lfloor\frac{n}{2}\rfloor$.
Turns out that, for a given sector $\dk$, the integral operator 
has a set of left eigenfunctions in the form
\begin{equation}
S^\lambda_q(v;n)=\mathbb{E}_h\int\dd u\ g^\lambda_q(u;n)\left[\frac{\cosh\left(\beta( u+ v+ h)\right)}{2\cosh(\beta u)\cosh( \beta v)}\right]^n
\left[1-\tanh^2\left(\beta( u+ v+ h)\right)\right]^q,
\label{sq}
\end{equation}
as can be inferred from  Eq. \eqref{psi-l} and can be directly verified. In the rest of the paper we will assume that the left and right eigenfunctions of the sector $\dk$ satisfy the normalization condition
\begin{equation}
\int \dd u \ S^\lambda_q(u;n)\, g^{\lambda'}_q(u;n) = \delta_{ \lambda\lambda'}
\label{normgs}
\end{equation}
derived from Eq. \eqref{normrho}.
We are now ready to take the $n\downarrow 0$ limit and discuss its non trivial aspects.
\subsection{The small $n$ limit}
\label{subsec:smalln}


\begin{figure}
\includegraphics[width=\textwidth]{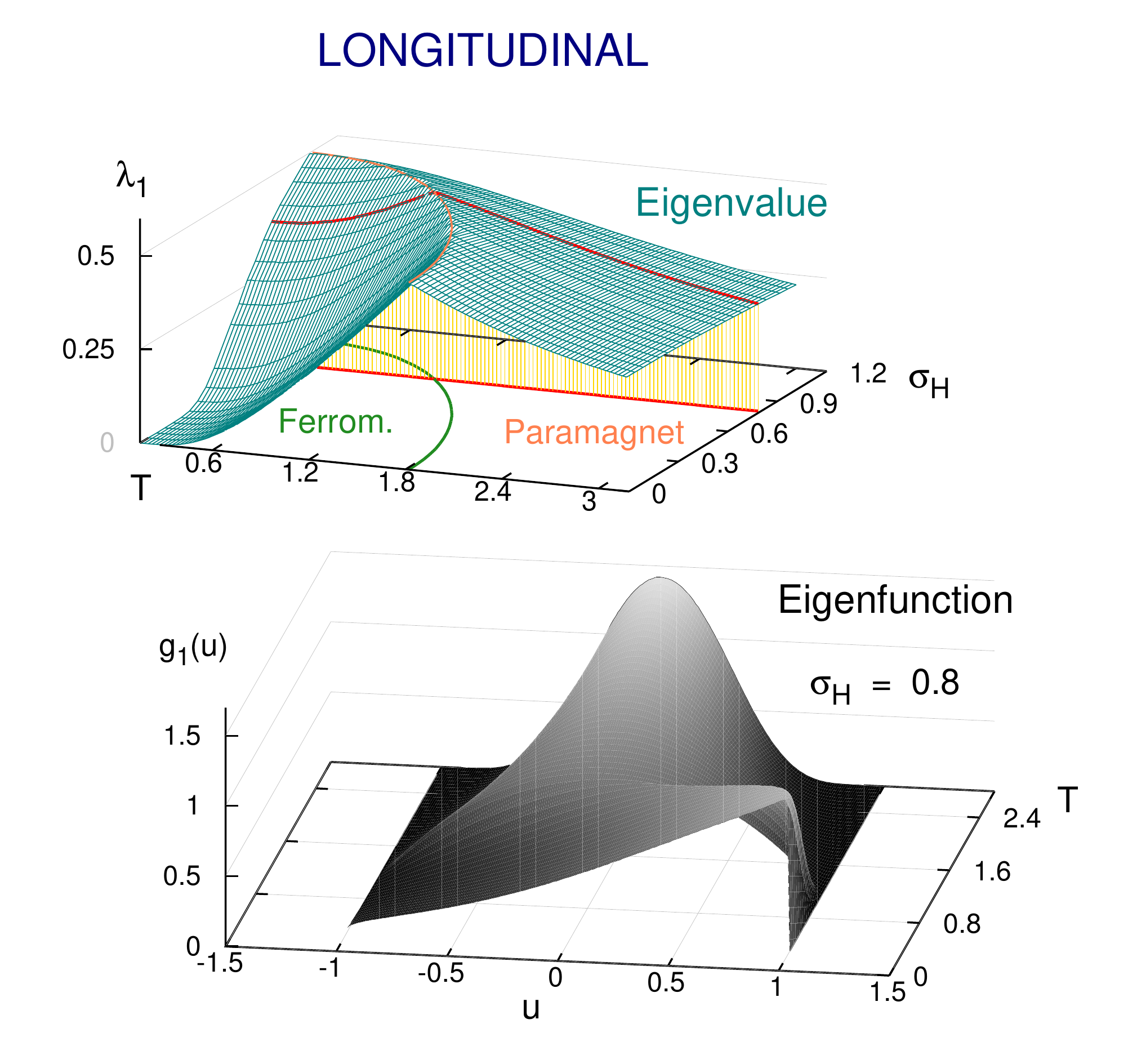}
\caption{($Top$) The leading eigenvalue $\lambda_1$ of the sector $\dw$ in the RFIM, as a function of the temperature and of the gaussian external field with variance $\sigma^2_H$. ($Bottom$) The corresponding right eigenfunction $g_1(u)$  at $\sigma_H=0.8$  . The random fields $h$ and $\hh$ are distributed as the cavity fields arriving on a chain embedded in a RRG with connectivity $z=3$, therefore the transition point is localized at $\lambda_1=\frac{1}{2}$.}
\label{img:eig1}
\end{figure}

\begin{figure}
\includegraphics[width=\textwidth]{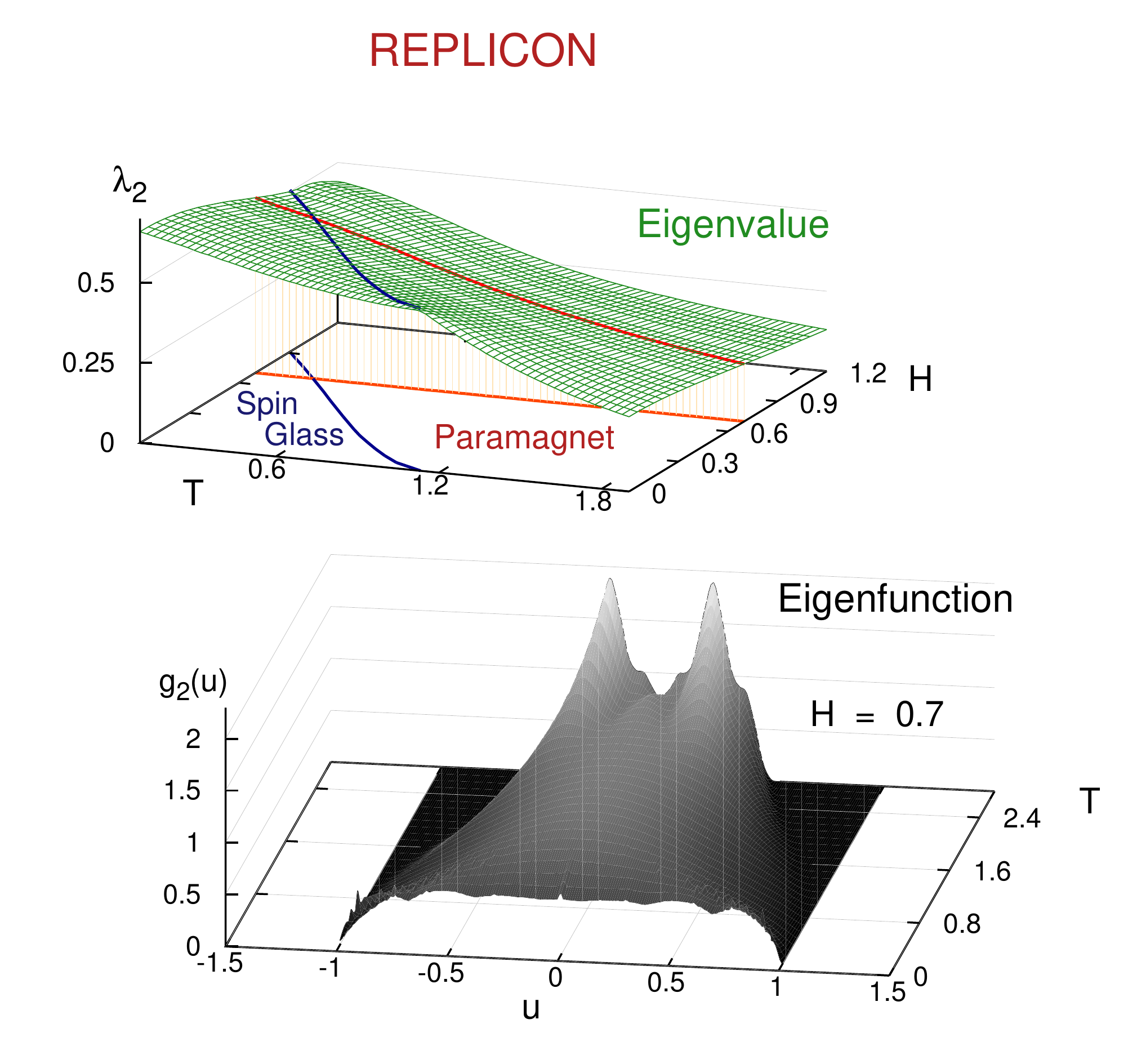}
\caption{($Top$) The leading eigenvalue $\lambda_2$ of the sector $\de$ in a $J=\pm 1$ spin-glass, as a function of the temperature and of the uniform external field $H$. The phase diagram is also shown in the $H-T$ plane. ($Bottom$) The corresponding right eigenfunctions $g_2(u)$  along the orange line of the top picture. The random fields $h$ and $\hh$ are distributed as the cavity fields arriving on a chain embedded in a RRG with connectivity $z=3$, therefore the transition point is localized at $\lambda_2=\frac{1}{2}$. }
\label{img:eig2}
\end{figure}

In the limit  $n\downarrow 0$ we obtain an infinite number of sectors $\dk$, $q=0,1,\ldots$, in a fashion that is characteristic to replicas computations. Setting $n=0$ in Eq. \eqref{eqeig-n} we obtain Eq. \eqref{eig-g}, which we rewrite for convenience: 

\begin{equation}
\begin{aligned}
\lambda\ g^\lambda_q(u)=\mathbb{E}_{J,h}\int \dd v\ \delta\left(u-\hat{u}(J,h+v)\right)\left(\frac{1}{\beta}\frac{\partial\hat{u}}{\partial{v}}\right)^q  g^\lambda_q(v)\ .
\label{eig-g-2}
\end{aligned}
\end{equation}
From now on we shall refer to $g^\lambda_q(v)$ as a solution of last equation and shall explicitly express the $n$ dependence for the solutions of \eqref{eqeig-n} at finite $n$. In Figure \ref{img:eig1} and Figure \ref{img:eig2} we show two examples of eigenvalues and eigenfunctions in the sector $\dw$ and $\de$ respectively.

For $q=0$ , i.e. in the sector $\dq$,   Eq. \eqref{eig-g-2} admits a unique maximum eigenvalue $\lambda=1$ by Perron-Frobenius theorem. The corresponding eigenfunction is the probability distribution of cavity biases, which we call $P(u)$ \cite{Montanari2009}. We have thus established a first connection between the cavity method and the RTM formalism, and we shall enforce this connection in Section \ref{sec:cavity}. The other eigenfunctions of $\dq$ are characterized by  $\int \dd u\ g^\lambda_0(u) = 0$ at $n=0$. It is convenient, to held compatibility with the normalization condition Eq. \eqref{normgs} as we will see, to impose a diverging scaling for all the eigenfunctions of $\dq$ except for the first one:
\begin{equation}
g_0^\lambda(u;n)\sim \frac{1}{\sqrt{n}}\left(g_0^\lambda(u)+n\, \tilde{g}_0^\lambda(u)\right).
\label{gon}
\end{equation}
The symbol $\sim$ denotes equivalence between the r.h.s. ad l.h.s. up to higher order correction in $n$, and $\tilde{g}^\lambda_0$ is the first correction to the leading order of the eigenfunction in $\dq$. Using Eq. \eqref{gon} for the right eigenfunctions and considering also the correction in $n$ to the eigenvalues, we can compute the left eigenfunctions of $\dq$ from Eq. \eqref{sq}. In fact we obtain at the leading order
\begin{equation}
S^\lambda_0(v;n)\sim\sqrt{n}\, S^\lambda_0(v) =\sqrt{n}\left[\mathrm{c}_\lambda+\mathbb{E}_h\int\dd u\  
g_0^\lambda(u)\,\log\left(\frac{\cosh\left(\beta( u+ v+ h)\right)}
{\cosh( \beta u)}\right)\right],
\label{s0}
\end{equation}
where $\mathrm{c}_\lambda$ is the normalization of the first order correction to the eigenfunction $g^\lambda_0$, that is
\begin{equation}
\mathrm{c}_\lambda \equiv \int\dd u\ \tilde{g}^\lambda_0(u)= \frac{1}{\lambda-1}\  \mathbb{E}_h\int \dd u\ g_0^\lambda(u)\,\log\left(\frac{\cosh\left(\beta( u+  h)\right)}
{\cosh( \beta u)}\right)\ .
\label{c-def}
\end{equation}
In all calculations involving the sector $\dq$ we will express the eigenvectors using Eqs. \eqref{gon} and \eqref{s0}, then proceed carefully to take the $n\downarrow 0$ limit.

To find an expression for the left eigenfunctions in the other sectors no such care is needed to take the $n\downarrow 0$ limit in Eq. \eqref{sq}, therefore we straightly obtain 
\begin{equation}
S^{\lambda}_q(v)=\mathbb{E}_h\int\dd u\ g^{\lambda}_q(u)
\left[1-\tanh^2\left(\beta( u+ v+ h)\right)\right]^q      \qquad\text{for } q\geq 1\ .
\label{sq-2}
\end{equation}
The degeneracy between $\dq$ and $\dw$ corresponds to the degeneracy between the Longitudinal and Anomalous eigenvalues in the Hessian of the Sherrington-Kirkpatrick model\cite{Thouless1978,Bray1979}.
The multiplicity of the eigenvalues in the two sectors,
$d_0=1$ and $d_1=n-1$, sum up to give an $O(n)$ contribution as should be expected, while from Eq. \eqref{dq} the other sectors have degeneracies of order $O(n)$ without the need of further elisions. Therefore it is convenient to define
\begin{equation}
 \hat{d}_q= 
 \begin{cases}
  1            &\text{for } q=1\ , \\
\lim_{n\to 0}\frac{d_q}{n} =  (-1)^{q+1} \frac{2q-1}{q(q-1)}           &\text{for } q\geq 2\ . \\
 \end{cases}
 \label{dhat} 
\end{equation} 

The first eigenvalue of $\dq$ requires separate considerations. We define the coefficient $f_0$ from its $n$ expansion:
\begin{equation}
\lambda(n)\sim 1-\beta f_0 n.
\end{equation}
As we already noted, the cavity messages distribution $P(u)$ is the eigenvector associated to the largest eigenvalue of the sector $\dq$ for $n=0$. The corresponding left eigenvalue is $S(u)\equiv 1$. In Section \ref{subsec:chains} we shall see that $f_0$ is the intensive free energy of a chain. From Eq. \eqref{eq:deltamu} we obtain
\begin{equation}
-\beta f_0 =\mathbb{E}_{J,h}\int \dd v\ \log\left[ Z(J,h,v)\right]
P(v) \ .
\label{f0}
\end{equation}


\subsection{The degeneracy between the Longitudinal and the Anomalous sector}
\label{subsec:degeneracy}
A close inspection of the eigenvalue equation \eqref{eig-g-2} reveals a surprising relation between the sectors 
$\dq$ and $\dw$ at $n=0$.  
It can be shown, respectively deriving or integrating both members of Eq. \eqref{eig-g-2} for $q=1$ and $q=0$, that all the eigenfunctions of $\dw$ have a corresponding eigenfunction in $\dq$ with the same eigenvalue. On the other hand, all the eigenfunctions of $\dq$, except for the first one, i.e. the ones having zero sum, have a corresponding eigenfunction in $\dw$ with the same eigenvalue. We have thus established a degeneracy between the Longitudinal and the Anomalous sectors. The following relations hold:
\begin{equation}
g^\lambda_0(u) = \frac{1}{\beta} \partial_u\, g_1^\lambda(u);\qquad \frac{1}{\beta} \partial_u\, S^\lambda_0(u) = -S_1^\lambda(u).
\label{rel01}
\end{equation}

Particular attention has to be taken in the limits involving these two sectors, keeping track of the $O(n)$ corrections both to eigenvalues and eigenvectors.
A double pole contribution to some observables, as we shall later see, stems from the first correction in $n$ to the paired  eigenvalues in $\dq$ and $\dw$. In fact if we define the eigenvalue shifts $\delta \lambda_0$ and $\delta\lambda_1$ by
\begin{align}
\lambda_0(n)&\sim\lambda + n\ \delta \lambda_0, \\
\lambda_1(n)&\sim\lambda + n\ \delta \lambda_1 ,
\end{align}
and consider the expansion to the first order in $n$ of the eigenvalue equation  \eqref{eqeig-n}  for $q=$, from standard perturbation theory we have
\begin{equation}
\delta\lambda_{q}=
\mathbb{E}_{J,h}\int\dd u \dd v\ S^\lambda_{q}(u) \log\left[ Z(J,h,v)\right]
\delta\left(u-\hat{u}(J,h+v)\right)
\left(\frac{1}{\beta}\frac{\partial\hat{u}}{\partial v}\right)^{q} g^\lambda_{q}(v) .
\label{eq:deltamu}
\end{equation}
The shift difference $\Delta_\lambda=\delta\lambda_0-\delta\lambda_1$ is the relevant quantity we are looking after, since it arises in the calculation of the free energies of closed chains and of the thermally disconnecter correlation function, see Section \ref{sec:applications}. Using Eq. \eqref{eq:deltamu} and the relation \eqref{rel01} between the eigenfunctions in the two sectors, we obtain the expression
\begin{equation}
\Delta_\lambda = -\mathbb{E}_{J,h}\int\dd u\dd v\  S^{\lambda}_0(u) \delta\left(u-\hat{u}(J,v+h)\right)\left[\tanh\left(\beta(v+h)\right)-\tanh(\beta v)\right]g^{\lambda}_1(v)
\label{Delta}
\end{equation}
If we call $(\bullet,\bullet)$ the scalar product in $L^2$ and define the kernel 
\begin{equation}
Q(u,v)=\mathbb{E}_{J,h}\,\delta\left(u-\hat{u}(J,v+h)\right)\left[\tanh\left(\beta(v+h)\right)-\tanh(\beta v)\right],
\end{equation}
then Eq. \eqref{Delta} can be rewritten as 
\begin{equation}
\Delta_\lambda = - (S_0^\lambda, Q\, g_1^\lambda).
\end{equation}
In the next Section we shall apply the formalism we have developed to the computation of some physically relevant quantities.

\section{Some Applications of the formalism}
\label{sec:applications}

\subsection{Free energy of chains}
\label{subsec:chains}
Let us first consider the average free energy of a closed chain of length $\ell$, each node receiving i.i.d.   random fields $h$, and call it $f_\ell^c$. If the chain considered is embedded in a locally tree-like graph, the random fields $h$ are distributed according to the cavity messages distribution on that graph ensemble. Since $\tr T_n^\ell$ is the replicated partition function of this system, the free energy is given by
\begin{equation}
-\beta f_\ell^c=\lim_{n\to 0} \partial_n \tr T_n^\ell\ ,
\end{equation}
where, thanks the orthonormal decomposition of $T_n$,  the trace can be written in the form
\begin{align}
\tr T_n^{\ell}=\sum_{q=0}^{\frac{n}{2}}d_q\sum_{\lambda\in\dk}\lambda^{\ell}\ .
\end{align}
In last equation the eigenvalue degeneracies $d_q$ are given in Eq. \eqref{dq}, and the eigenvalues $\lambda$ depends implicitly on $n$. In the small $n$ limit the sum over $q$ can be extended to infinity.  The considerations over the eigenvalues' shifts and degeneracies of last Section lead to the final expression 
\begin{equation}
-\beta f_{\ell}^{c}=-\beta\ell  f_0
+\sum_{\lambda\in\dw}\Delta_\lambda\,\ell\,\lambda^{\ell-1}
+\sum_{q=1}^{\infty}\hat{d}_q
\sum_{\lambda\in\dk} \lambda^{\ell}\ .
\label{f-clos}
\end{equation}
The coefficients $\hat{d}_q$ are given by Eq. \eqref{dhat}, the shift differences $\Delta_\lambda$ given by Eq. \eqref{Delta} and an expression for the intensive free energy $f_0$ is found in Eq. \eqref{f0}. We notice that all the quantities entering Eq. \eqref{f-clos} can be expressed in terms of the eigenvalues and eigenfunctions of Eq. \eqref{eig-g-2}. 

The computation of the average free energy of open chains is a little more involved.
In the definition of open chains, we allow the spins at the extremities to receive a random field $\hh$ that could have a distribution different from the one of the fields acting on the internal spins of the chain. We introduce this relaxation of the model in order to apply our formalism to the case of open chains embedded in a generic tree-like random graph. 

It is convenient to define the replicated partition function of an open chain of length $\ell$, conditioned on the configuration of the replicated spins at its extrema in the following way: starting from $T_n^\ell$, we remove the field $h$ on the right and substitute it with a field $\hh$, then we add the other field $\hh$ on the left (see Figure \ref{fig:ttt}). Therefore we define
\begin{equation}
\tilt(\s,\f)\equiv \rho_{\hh}(\s) \,T_n^{\ell}(\s,\f)\,\rho^{-1}_{h}(\f)\rho_{\hh}(\f),
\label{ttilde}
\end{equation}
where, with a little abuse of notation, the vector $\rho_{\hh}$ is defined by
 \begin{equation}
\rho_{\hh}(\s)\equiv \mathbb{E}_{\hh}\, e^{\beta \hh\sum_a  \s^a}.
\end{equation}
By definition the matrix $\tilt$ is symmetric (see Figure \ref{fig:ttt} for a pictorial representation). From Eq. \eqref{ttilde} and Eq. \eqref{spectr-dec} we obtain the spectral decomposition
\begin{equation}
\tilt(\s,\f)=\sum_{q=0}^{\lfloor\frac{n}{2}\rfloor}\sum_{\lambda\in\dk}\lambda^\ell\ \rho_{\hh}(\s)\rho_q^{\lambda}(\s)\rho_{\hh}(\f)\rho_q^{\lambda}(\f)
\sum_{
\substack{
 a_1<\cdots<a_q  \\
b_1<\cdots<b_q  
}}
Q_{a_1\dots a_q;b_1\dots b_q}\,
\s^{a_1}\dots\s^{a_q}\f^{b_1}\dots\f^{b_q}.
\label{tilde-t}
\end{equation}  
 
The average free energy of an open chain of length $\ell$ is then given by
\begin{equation}
-\beta f_\ell^o=\lim_{n\to 0} \partial_n\ \sum_{\s,\f} \tilt(\s,\f)\ .
\end{equation}
From Eq. \eqref{tilde-t} it easy to see that only the $\dq$ sector of $\tilt$ contributes to last equation. 

 A different behaviour characterize the terms corresponding to the leading eigenvalue at $n=0$ (the cavity one) from the others. As in the case of the closed chain, the extensive contribution to the free energy comes from the leading eigenvalue of $\dq$, $\lambda\sim 1-n\,\beta f_0 $. An $O(1)$ contribution comes from the leading eigenfunction $g^\lambda_0(u;n) = P(u)+O(n)$, while each other eigenvalue, the ones degenerate with $\dq$, gives an exponential term.
Therefore, after a careful treatment of the small $n$ limit, we arrive to the expression
\begin{equation}
\begin{aligned}
-\beta f^o_{\ell}=& -\ell\beta f_0+\mathbb{E}_{\hh}\int \dd u\ P(u)\,2\log\cosh\left(\beta( u+\hh)\right)-\mathbb{E}_{h}\int \dd u\dd v\ P(u)P(v)\log\cosh\left(\beta( u+v+h)\right)\\
&+\log 2 +\sum_{\lambda\in D^{(1)}} a_{\lambda,0}^2\,\lambda^\ell \ , 
\end{aligned}
\label{fop-2}
\end{equation}
with
\begin{equation}
a_{\lambda,0}=\frac{1}{\lambda-1}\  \mathbb{E}_h\int \dd u\ g_0^\lambda(u)\,\log\left[\frac{\cosh\left(\beta(u+h)\right)}{\cosh(\beta u)}\right]+\mathbb{E}_{\hh}\int \dd u\ g_0^\lambda(u)\,\log\left[\frac{\cosh(\beta(u+\hh))}{\cosh(\beta u)}\right]
\label{b-lambda}
\end{equation}
In Eq. \eqref{fop-2} it is clearly expressed at the order $O(1)$ in $\ell$ the free energy shift, with respect to the free energy of a closed chain, due to the addition of two extremal spins and the removal of an internal one.

The coefficients $a_{\lambda,0}$ are strictly related to the left eigenfunctions $S^\lambda_0$ defined in Eq. \eqref{s0}. In fact if the random field at the extremities of the chain are distributed as the one on the internal spins, i.e. $\hh \,{\buildrel d \over =}\, h$ as in the case of a chain embedded in a Poissonian random graph, then $a_{\lambda,0}=S^\lambda_0(0)$. More generally if a probability distribution  $G^{(0)}(u)$ exists such that
\begin{equation}
\tilde{P}(\hh) = \mathbb{E}_{h} \int \dd u \ G^{(0)}(u)\ \delta(\hh -(u+h))
\label{init}
\end{equation}
holds, then Eq. \eqref{b-lambda} can be written in the compact form $
a_{\lambda,0} = (S_0^\lambda,G^{(0)})$.
Obviously if  $\hh \,{\buildrel d \over =}\, h$ we have $G^{(0)}(u)=\delta(u)$. For a chain embedded in a random regular graphs ensemble instead,  $G^{(0)}(u)$ is given by the distribution of cavity biases $P_{cav}(u)$\cite{Montanari2009}, which corresponds to the first eigenvector of the Longitudinal Sector. Therefore in the random regular graph ensemble  $(S_0^\lambda,G^{(0)})=0$ and no exponential decays are present in the expression \eqref{fop-2} for the free energy of open chains.

\subsection{Correlation functions}
\label{subsec-corr}
We take advantage of the spectral representation of the RTM to find some analytical expressions for the  two-point correlation functions. We consider two spins, $\s_0$ and $\s_\ell$, at distance $\ell$ along a chain. As in the previous paragraph, we admit the possibility for the chain to be embedded in a locally tree-like graph, therefore the random fields $\hh$ acting on $\s_0$ and $\s_\ell$ can be distributed differently from the fields $h$ on the internal spin of the chain.  The decomposition of $\tilt(\s,\f)$ in Eq. \eqref{tilde-t} can be exploited  to obtain the correlation functions. In fact contracting $\tilde{T_n}^{\ell}(\s,\f)$ with two spins having the same replica index constrains them to be in the same thermal state, as in 
$\overline{\langle \s_0 \s_\ell\rangle} 
= \lim_{n\to 0}\sum_{\s,\f} \s^1 
\,\tilt(\s,\f)\, \f^1 $.
Choosing different replica indexes instead corresponds to choosing different thermal states, e.g. 
$\overline{\langle \s_0 \rangle\langle\s_\ell\rangle} 
= \lim_{n\to 0}\sum_{\s,\f} \s^1 
\,\tilt(\s,\f)\, \f^2 $. Generalizing this considerations is easy to obtain
\begin{equation}
\overline{\langle \s_0\s_{\ell}\rangle^k}=\ \lim_{n\to0}\ \sum_{\s,\f}\s^1\ldots \s^k\ \tilt(\s,\f)\ \f^1\ldots \f^k.
\label{nonconn-q}
\end{equation}
Since vectors of the form $\s^1\ldots \s^k$ have non-zero projections in $\dk$ only for $q\leq k$, only these sectors of the spectral representation of $\tilt$ contribute to Eq. \eqref{nonconn-q}. The expression for $\overline{\langle \s_0\s_{\ell}\rangle^k}$ is quite complicated and it involves also the correction for small $n$ to the eigenfunction of $\dq$ and $\dw$, as in the case of the thermally disconnected correlation function we shall later see. Therefore, since this kind of correlation function has little physical relevance, we won't report its expression in terms of the transfer matrix eigenvalues and eigenfunctions. 

Far more interesting from the physical viewpoint are the connected correlation functions. The ferromagnetic connected correlation functions can be expressed as  $
\overline{\langle \s_0\s_{\ell}\rangle_{\mathrm{c}}}=\ \lim_{n\to0}\ \frac{1}{2}\sum_{\s,\f}\left(\s^1-\s^2\right)\, \tilt(\s,\f)\, \left(\f^1-\f^2\right)$, as one can rapidly check, and this expression can be easily generalized to

\begin{equation}
\overline{\langle \s_0\s_{\ell}\rangle_{\mathrm{c}}^k}=\ \lim_{n\to0}\ \frac{1}{2^k}\sum_{\s,\f}\left(\s^1-\s^2\right)\ldots\left(\s^{2k-1}-\s^{2k}\right)\ \tilt(\s,\f)\ \left(\f^1-\f^2\right)\ldots\left(\f^{2k-1}-\f^{2k}\right)\ .
\end{equation}
t 
It is worth noticing that the vector $v= \left(\s^1-\s^2\right)\ldots\left(\s^{2k-1}-\s^{2k}\right)
$ belongs to the subspace $D^{(k)}$, therefore we can choose a basis for the spectral representation of  $\tilt$ such that all but one vectors are orthogonal to $v$. This leads to the following compact expression for the connected correlation functions:

\begin{equation}
\overline{\langle \s_0\s_{\ell}\rangle_{\mathrm{c}}^k }=\sum_{\lambda\in D^{(k)}}a^2_{\lambda, k}\ \lambda^{\ell}
\ ,
\label{corr-conn}
\end{equation}
with the coefficients $a_{\lambda, k}$ given by 
\begin{equation}
a_{\lambda,k}=\mathbb{E}_{\hh}\int\dd u\ g^\lambda_k(u)\ 
\left[1-\tanh^2\left(\beta( u+ \hh)\right)\right]^k\ . 
\label{alk}
\end{equation}
As in the case of the coefficient $a_{\lambda,0}$ defined in Eq. \eqref{b-lambda}, if a solution $G^{(0)}$ of \eqref{init} exist then $a_{\lambda,k}$ is simply given by the projection of $G^{(0)}$ on $S_k^\lambda$, that is $a_{\lambda,k}=(S_k^\lambda,G^{(0)})$.

Equation $\eqref{corr-conn}$ allows us to easily compute the susceptibilities $\chi_k=\lim_{N\to\infty} \frac{1}{N}\sum_{i,j} \overline{\langle \s_i \s_j\rangle_c^k}$ in a random graph with mean degree and mean residual degree $z_0$ and $z$ respectively, in fact in thermodynamic limit we have
\begin{equation}
\begin{aligned}
\chi_k &= \overline{(1-m^2)^k} +\sum_{\ell=1}^{\infty} z_0 z^{\ell-1} \,\overline{\langle \s_0\s_{\ell}\rangle_c^k} \\
&=\overline{(1-m^2)^k} +z_0\sum_{\lambda\in D^{(k)}} a^2_{\lambda, k}\,\frac{\lambda}{1-z\lambda}
\end{aligned}
\end{equation}
At a transition point the largest eigenvalue of one of the sectors $\dk$ reaches the value $\frac{1}{z}$ and the corresponding susceptibility diverges. Assuming a smooth behavior for the eigenvalue in the high temperature region before the transition, $\lambda(T) = \frac{1}{z}+O(T-T_c)$ for  $T\to T_c^+$, we obtain the mean-field critical exponent $\gamma=1$. 

The computation of the thermal disconnected correlation function $\overline{\langle\s_0\rangle\langle \s_\ell\rangle}$, relevant to the RFIM transition, is more complicated, since it involves the sub-leading corrections in $n$ to the eigenvectors of $T_n$. Great care has to be taken in the limit $\lim_{n\to 0}\sum_{\s,\f} \s^1 
\,\tilt(\s,\f)\, \f^2 =\overline{\langle\s_0\rangle\langle \s_\ell\rangle}$. As in Eq. \eqref{gon}, let us call $\tilde{g}^\lambda_0(u)$ the correction to the eigenfunction $g^\lambda_0(u)$. We denote with $\langle\bullet\rangle_q$ the expectation over $\tilt(\s,\f)$ restricted to the sector $\dk$. Than in $\dq$ we obtain
\begin{equation}
\begin{aligned}
\langle\s^1 \f^2\rangle_0 \sim\ & \overline{\langle\s_\infty \rangle}^2+\sum_{\lambda\in \dw}\frac{1}{n}a^2_{\lambda,1}\, \lambda^{\ell}+a^2_{\lambda,1}\,\ell\,\delta\lambda_0\,\lambda^{\ell-1}\\
&-2 a_{\lambda,1}\,\lambda^{\ell}\left[\int\dd u\ \tilde{g}^\lambda_0(u)\tanh(\beta(u+\hh))+\int \dd u\ g^\lambda_0(u) \tanh(\beta(u+\hh))\log\frac{\cosh(\beta(u+\hh))}{\cosh(\beta u)}\right],
\end{aligned}
\label{disc-contr-0}
\end{equation} 
where the contribution $\overline{\langle\s_\infty \rangle}$ comes from the cavity eigenvector and is the average magnetization of a spin at the end of an infinite chain.

Similarly, if we define $\tilde{g}^\lambda_1(u)$ by $g^\lambda_1(u;n)\sim g^\lambda_1(u)+n\, \tilde{g}^\lambda_1(u)$, in the sector $\dw$ we have
\begin{equation}
\begin{aligned}
\langle\s^1 \f^2\rangle_1 \sim\ &\sum_{\lambda\in \dw}\frac{-1}{n}a^2_{\lambda,1}\, \lambda^{\ell}-a^2_{\lambda,1}\,\ell\,\delta\lambda_1\,\lambda^{\ell-1}\\
&-2 a_{\lambda,1}\,\lambda^{\ell}\left[\int\dd u\ \tilde{g}^\lambda_1(u)\left(1-\tanh^2(\beta(u+\hh))\right)+\int \dd u\  g^\lambda_1(u)\left(1-\tanh^2(\beta(u+\hh))\right)\log\frac{\cosh(\beta(u+\hh))}{\cosh(\beta u)}\right]
\end{aligned}
\label{disc-contr-1}
\end{equation} 
Summing the two contributions, the final result for the disconnected correlation function is
\begin{equation}
\overline{\langle\s_0\rangle\langle \s_\ell\rangle}= \overline{\langle\s_\infty \rangle}^2+
\sum_{\lambda\in \dw}\Delta_\lambda\, a^2_{\lambda,1}\,\ell\,\lambda^{\ell-1}
+\alpha_\lambda\,\lambda^{\ell}.
\label{c-disc2}
\end{equation}
Therefore each eigenvalue of the Anomalous sector contributes to $\overline{\langle\s_0\rangle\langle \s_\ell\rangle}$  with an simple exponential term and with a term that leads to a double pole behaviour in the associated susceptibility, with coefficients $\Delta_\lambda$ given in Eq. \eqref{Delta}. The coefficients $\alpha_\lambda$ of the exponential decays instead are given by
\begin{equation}
\begin{aligned}
\alpha_\lambda=&\ 2\, a_{\lambda,1}\bigg[\int \dd u\  g^\lambda_1(u)\tanh(\beta(u+\hh))\left(\tanh(\beta(u+\hh))-\tanh(\beta u)\right)\\
&-\int\dd u\ \tilde{g}^\lambda_1(u)\left(1-\tanh^2(\beta(u+\hh))\right)
-\int\dd u\ \tilde{g}^\lambda_0(u)\tanh(\beta(u+\hh))\bigg]
\end{aligned}
\end{equation}

Since the magnetization of a spin conditioned to be to be the extremity of a chain of size $\ell$ is given by 
\begin{equation}
\overline{\langle\sigma_\ell\rangle}=\lim_{n\to 0}\sum_{\s,\f} \s^1 
\,\tilt(\s,\f)=\overline{\langle\s_\infty \rangle} -\sum_{\lambda\in\dw}a_{\lambda,1}\, a_{\lambda,0}\, \lambda^\ell,
\end{equation}
if we call $\Lambda$ the highest eigenvalue of the sector $\dw$, the most relevant contributions to the thermally-disconnected disorder-connected correlation function is given by
\begin{equation}
\overline{\langle\s_0\rangle\langle \s_\ell\rangle}- \overline{\langle\s_0 \rangle}\,\overline{\langle\s_\ell \rangle} =
\Delta_\Lambda\, a^2_{\Lambda,1}\,\ell\,\Lambda^{\ell-1}
+\left(\alpha_\Lambda -2 \overline{\langle\s_\infty \rangle }\, a_{\Lambda,1}\, a_{\Lambda,0}\right)\Lambda^{\ell}+o(\Lambda^\ell) \qquad \text{for } \ell\to+\infty.
\label{c-disc3}
\end{equation}
We notice that, while the coefficient of the exponential term is quite hard to compute, the coefficient $a^2_{\Lambda,1} \Delta{_\Lambda}$, which regulates the leading behaviour, has a much simpler expression given in Eq. \eqref{Delta} and Eq. \eqref{alk}. From Eq. \ref{c-disc3} turns out that near a ferromagnetic transition point, i.e. $\Lambda = \frac{1}{z}$, as long as $\Delta_{\Lambda}$ is not zero, the leading behavior of the disconnected susceptibility 
$\chi_{disc}=\sum_{i,j}\overline{\langle\s_0\rangle\langle \s_\ell\rangle}- \overline{\langle\s_0 \rangle}\,\overline{\langle\s_\ell \rangle}$ reads
\begin{equation}
\chi_{disc}\simeq z _0\, \Delta_{\Lambda}\,a^2_{\Lambda,1} \frac{1}{(1-z\Lambda)^2}.
\end{equation}
The expected double-pole behavior of the disconnected susceptibility is thus recovered.

\section{Cavity derivation}
 \label{sec:cavity}

In this section we present the derivation of several of the results of last Section using a probabilistic approach, in the same spirit of the usual cavity method calculations \cite{Parisi1987, Montanari2009}. While this approach is more physically intuitive than the RTM formalism, it requires the set up of an ad-hoc recursion rule for each observable. Noticeably we could not recover Eq. \eqref{f-clos} for the free energy of closed chains.

\subsection{Open chains}
\label{sub-cav-open}
We want to study the statistical properties of a random Ising open chain without the use of replicas. We start with an asymmetric chain of length $\ell$, whose random partition function we denote with $Z_\ell$, constructed iteratively according to the following procedure: $Z_0$ is the partition function of a single spin receiving a random field $u_0$, i.e. $Z_0=2\cosh(\beta u_0)$; at the $i$-th step of the construction we add a spin $\s_i$, a random coupling $J_i$ between $\s_i$ and $\s_{i-1}$ and a random field  $h_{i-1}$ on $\s_{i-1}$; the random variable $Z_\ell$ is the partition function of the system obtained after the $\ell$-th step of the procedure. Note that the last spin added to the chain has no external fields acting on it. The following distributional identity can be easily derived:
\begin{equation}
Z_{\ell+1}=
\frac{2\cosh(\beta J_\ell)\,\cosh(\beta(u_\ell+h_\ell))}{\cosh(\beta u_\ell)}\times Z_\ell \equiv  Z(J_\ell,h_\ell,u_\ell)
\times Z_{\ell}. 
\label{ric-zl}
\end{equation}
It is convenient to introduce the quantity $\overline{Z^n_{\ell}(u)}\equiv\overline{\delta(u-u_\ell)\,Z^n_{\ell}}$, which corresponds to the expectation of $Z_\ell^n$ along with the indicator function of the event $u_\ell=u$. Here $n$ is an arbitrary  chosen positive real number, the symbol being chosen to stress the analogy with the replica formalism where the quantity $n$ (integer in this case) is the number of replicated systems. Using this definition from Eq. \eqref{ric-zl} follows readily
\begin{equation}
\begin{aligned}
\overline{Z^n_{\ell+1}(u)} = \mathbb{E}_{J,h}\int \dd v \ \delta\left(u-\hat{u}(J,v+h)\right)\, Z^n(J,h,v)\ 
\overline{Z^n_{\ell}(v)},
\end{aligned}
\label{ricor-znl}
\end{equation}
where $\tilde{u}(J,x)=\frac{1}{\beta}\atanh(\tanh(\beta J)\tanh(\beta x))$ is the usual message passing rule.
The integral operator of Eq. \eqref{ricor-znl} is the same we found in the RTM formalism in Eq. \eqref{eqeig-n} for the sector $\dq$, therefore we can make use of the spectral analysis result from those paragraphs, in particular of the completeness relation 
\begin{equation}
\mathbb{E}_{J,h}\ \delta\left(u-\hat{u}(J,v+h)\right)\, Z^n(J,h,v) = \sum_{\lambda\in\dq}\lambda(n)\ g_0^\lambda(u;n)\ S_0^\lambda(v;n),
\label{kern-d0-decomp}
\end{equation}
between left and right eigenvectors. The definition of the left eigenfunctions of $\dq$ was already given in Eq. \eqref{sq}, but we rewrite it for convenience:
\begin{equation}
S^\lambda_0(v;n)=\mathbb{E}_h\int\dd u\ g^\lambda_0(u;n)\left[\frac{\cosh\left(\beta( u+ v+ h)\right)}{2\cosh(\beta u)\cosh( \beta v)}\right]^n.
\label{cav-so}
\end{equation}

Let us define another random partition function, $Z_\ell(u;x)$, obtained from $Z_\ell(u)$ conditioning on the value of the message $u_0$ on the first spin, that is  
$Z_\ell(u;x)\, =\, Z_\ell(u) | (u_0=x)$. Since also $\overline{Z^n_\ell(u;x)}$ as a function of $u$ obeys equation \eqref{ricor-znl}, using the decomposition Eq. \eqref{kern-d0-decomp} and the initial condition $Z_0 = 2\cosh(\beta u_0)$ we arrive to the important result
\begin{equation}
\overline{Z^n_{\ell}(u;x)}=\sum_{\lambda\in\dq}\lambda^\ell(n)\, g_0^\lambda(u;n)\, S_0^\lambda(x;n)\, [2\cosh(\beta x)]^n.
\label{znluv} 
\end{equation}
Using last equation it is easy to compute any moment $\overline{Z^n_\ell}$, $n$ not necessarily integer, of the partition function of a random  asymmetric Ising chain of length $\ell$. More interesting is the computation of the properties of a symmetric Ising open chain, the one considered in Section \ref{subsec:chains}, which receives on each extremity an external field distributed according to a certain probability distribution $\tilde{P}(\hh)$. As already stated, this is definition stems from the need to cover the important case of a chain embedded in a locally tree-like graph. Let us call $Z_{\ell,o}$ the random partition function of this open chain. It is related to the  random partition function  $Z_\ell$ of the asymmetric open chain by
\begin{equation}
Z_{\ell,o}=\frac{\cosh(\beta(u_\ell+\hh_\ell))}{\cosh(\beta u_\ell)} \times Z_{\ell-1}(u_\ell;u_1)\times \frac{2\cosh(\beta J_0)\cosh(\beta\hh_0)}{\cosh(\beta u_1)}
\label{cav-zop}
\end{equation}
where $u_1$ is distributed as $\tilde{u}(J_0,\hh_0)$. 
From Eq. \eqref{cav-zop} along with Eq. \eqref{znluv} and Eq. \eqref{cav-so}, we derive the main result of this paragraph: 
\begin{equation}
\overline{Z^ n_{\ell,o}} = \sum_{\lambda\in\dq}\lambda^\ell(n)\ a^2_{\lambda,0}(n).
\label{znlo}
\end{equation}
where $a_{\lambda,0}(n)$ is defined by
\begin{equation}
\begin{aligned}
a_{\lambda,0}(n) \equiv \mathbb{E}_{\hh}\int \dd u\ \left[\frac{\cosh(\beta(u+\hh))}{\cosh(\beta u)}\right]^n g_0^\lambda(u;n). 
\end{aligned}
\label{a0n}
\end{equation}
In the RTM formalism of Section \ref{sec-decomp} and \ref{sec:applications}, last expression could be derived from $\tilt$ defined in Eq. \eqref{tilde-t} by analytic continuation of $\overline{Z^ n_{\ell,o}}=\sum_{\s,\f}\tilt(\s,\f)$ to non-integer $n$. 

The average free energy of an open chain of length $\ell$ can then be obtained by
\begin{equation}
-\beta f_{\ell}^{o}= \lim_{n\to 0} \partial_n\ \overline{Z^ n_{\ell,o}}.
\end{equation}
The computation involves computing the order $n$ of all the quantities present in Eq. \eqref{znlo}, as it was done in Section \eqref{subsec:chains}. 
In this paragraph however, without any use of replicas, we gave a purely probabilistic argument valid for any real value of $n$. We refer therefore to Section \eqref{subsec:chains} for the successive step of the computation of $f_{\ell}^{o}$, leading to the final result Eq. \eqref{fop-2}. Notice that in the notation of that paragraph  $a_{\lambda,0}$ is related to $a_{\lambda,0}(n)$ defined in Eq. \eqref{a0n} by $a_{\lambda,0}(n)\sim \sqrt{n}\ a_{\lambda,0}$.

The expression \eqref{fop-2} for $f_{\ell}^{o}$ could also be obtained by a different approach that does not involve any limit $n\downarrow 0$ but is technically more difficult. We define the function $\varphi^{(\ell)}(u)$ by
\begin{equation}
\varphi^{(\ell)}(u) \equiv \overline{\delta(u-u_\ell)\log Z_\ell},
\end{equation}
and observe that given the distribution of the cavity message at distance $\ell$ along the chain, $u_\ell$, which we call  $G_0^{(\ell)}(u)$, it obeys the iterative rule
\begin{equation}
\begin{aligned}
\varphi^{(\ell+1)}(u)\ =&\ \mathbb{E}_{J,h}\int \dd v\ \delta\left(u-\hat{u}(J,h+v)\right)\,\varphi^{(\ell)}(v)\\
&+\mathbb{E}_{J,h}\int \dd v\ \delta\left(u-\hat{u}(J,h+v)\right)\log\left[\frac{2\cosh(\beta J)\cosh(\beta(v+h))}{\cosh(\beta v)}\right]
\,G_0^{(\ell)}(v)
\end{aligned}
\label{ricor-varphi}
\end{equation}
Last equation can be solved decomposing $\varphi^{(\ell)}(u)$ and $G_0^{(\ell)}(v)$ along the eigenfunctions of $\dq$ at $n=0$, then $\varphi^{(\ell)}(u)$ can be used to obtain $f_{\ell}^{o}$.

\subsection{Connected correlation functions}
\label{subsec:cav-conn}
Let us derive the eigenvalue equation \eqref{eig-g} and the expression for the connected correlation functions Eq. \eqref{cq}, without making any use of replicas. Here we consider straightly the random open chain with partition function $Z_{\ell,o}$, characterized by independent random external field distributes a $h$ on the internal spins and as $\hh$ on the extremities. The connected correlation function $\langle \s_0\s_{\ell}\rangle_c=\frac{1}{\beta}\frac{\partial \langle \s_{\ell}\rangle}{\partial H_0}$, where $H_0$ is an auxiliary field acting on $\s_0$, can be expressed as a function of the message $u_\ell$, coming through the chain to the spin $\s_\ell$, and its derivative with respect to $H_0$. In fact we have
\begin{equation}
\langle \s_0\s_{\ell}\rangle_c
=\left(1-\tanh^2(\beta(\hh_\ell +u_\ell))\right) \frac{\partial u_\ell}{\partial H_0},
\label{conn-cav}
\end{equation}
where $\hh_\ell$, as usual, is the random effective field acting on $\s_\ell$ and coming eventually from the rest of the graph. Let us define the random variable $X_\ell$ by $X_\ell\equiv \frac{\partial u_\ell}{\partial H_0}$. The average over disorder of Eq. \eqref{conn-cav} and its moments  $\overline{\langle \s_0\s_{\ell}\rangle_c^q}$  can then be computed once we know the joint law of the random variables $u_\ell$ and $X_\ell$ , which we call $P_\ell(u,X)$.
Since $X_\ell$ obeys the chain rule  $X_{\ell+1}= \frac{\partial u_{\ell+1}}{\partial u_{\ell}} X_{\ell}$ the recursion rule for $P_\ell$ reads

\begin{equation}
P_{\ell+1}(u,X)\ =\ \mathbb{E}_{J,h}\int \dd v\, \dd Y\ \ \delta\left(X-\frac{\partial \hat{u}}{\partial v}\ Y\right)\delta\left(u-\hat{u}(J,h+v)\right)\, P_{\ell}(v,Y), 
\label{eq:recursionPC}
\end{equation}
where $\hat{u}$ is the message passing rule defined in Eq. \eqref{hatu}. 
From last expression it turns out we can write an iteration rule  for the momenta of $X_\ell$ at fixed $u_\ell$,
\begin{equation}
G_q^{(\ell)}(u)=\int \dd X\ P_{\ell}(u,X)\ X^q\ ,
\label{gql}
\end{equation}
which reads
\begin{equation}
G_q^{(\ell+1)}(u)\ =\ \mathbb{E}_{J,h}\int \dd v\ \delta\left(u-\hat{u}(J,h+v)\right)\left(\frac{\partial \hat{u}}{\partial v}\right)^q\, G_q^{(\ell)}(v).
\label{rec-gq}
\end{equation}
Equations \eqref{eq:recursionPC} and \eqref{rec-gq} with $q=2$ have been recently introduced in literature\cite{Janzen2010a} in order to derive an analytical expression for the spin-glass susceptibility. 

We note that the knowledge of the maximum eigenvalue of the integral operator of Eq. \eqref{rec-gq} for a generic $q$ allows one to reconstruct the full distribution of the connected correlation function at large distance \cite{Morone13}.

From last equation it is clear the relation of $G_q^{(\ell)}$ with the eigenfunctions $g_q^{\lambda}$ of Eq. \eqref{eig-g}. In fact, decomposing $G_q^{(\ell)}(u)$ along the eigenfunctions of $\dq$, projecting Eq. \eqref{rec-gq} on the left eigenvectors  $S^\lambda_q(u)$ and with some computations analogue to the ones leading from Eq. \eqref{znlo} to Eq. \eqref{a0n}, we arrive to
\begin{equation}
G_q^{(\ell)}(u)\ =\sum_{\lambda\in\dk}a_{\lambda,q}\, \lambda^\ell\, g_q^{\lambda}(u),
\label{Glq}
\end{equation}
where $a_{\lambda,q}$ is defined in Eq. \eqref{alk}. Equation \eqref{Glq}, along with Eq. \eqref{conn-cav}, gives the expression \eqref{corr-conn} obtained with the RTM formalism for the connected correlation functions.

Following the lead of the previous paragraph, we can extend the above derivation  to compute the disorder averages $\langle \s_0\s_{\ell}\rangle^q_c$ along with an arbitrary power of $Z_{\ell,o}$. The generalization of Eq. \eqref{rec-gq} in fact becomes
\begin{equation}
G_q^{(\ell+1)}(u;n)\ =\ \mathbb{E}_{J,h}\int \dd v\ \delta\left(u-\hat{u}(J,h+v)\right)\left(\frac{\partial \hat{u}}{\partial v}\right)^q Z(J,h,v)\, G_q^{(\ell)}(v;n),
\label{rec-gqn}
\end{equation}
and Eq. \eqref{Glq} generalizes trivially as well. The final result is
\begin{equation}
\overline{\langle \s_0\s_{\ell}\rangle^q_c\ Z^n_{\ell,o}}
=\sum_{\lambda\in\dk} \lambda^\ell(n)\ a^2_{\lambda,q}(n),
\label{conn-cav-zn}
\end{equation}
which extrapolates smoothly to the result we obtained for $n=0$, i.e. Eq. \eqref{corr-conn}. In last equation the coefficients $a_{\lambda,q}(n)$ are defined by 

\begin{equation}
\begin{aligned}
a_{\lambda,q}(n) \equiv \mathbb{E}_{\hh}\int \dd u\ \left[\frac{\cosh(\beta(u+\hh))}{\cosh(\beta u)}\right]^n
\left[1-\tanh^2(\beta( u+ \hh))\right]^q g_q^\lambda(u;n),
\end{aligned}
\label{aqn}
\end{equation}
such that $a_{\lambda,q}(0)=a_{\lambda,q}$.

\subsection{The disconnected correlation function}
\label{sub-cav-disc}
The computations of the thermally disconnected  correlation function $\overline{\langle \s_0\rangle\langle\s_{\ell}\rangle}$  is straightforward  once we use the results we obtained in the two preceding  paragraphs. In fact, calling $H_0$ and $H_\ell$ two auxiliary fields we add to the first and the last spin of the chain respectively and set to zero after the computation, the following relation holds: 
\begin{equation}
\frac{\partial}{\partial H_0}\frac{\partial}{\partial H_\ell}\overline{Z^n_{\ell,o}} = 
n\, \overline{\langle \s_0\s_{\ell}\rangle_c \ Z^n_{\ell,o}} + n^2\,\overline{\langle \s_0\rangle\langle\s_{\ell}\rangle\ Z^n_{\ell,o}}.
\end{equation}
Using Eqs. \eqref{znlo} and \eqref{conn-cav-zn} last expression leads to the main result of this paragraph, that is
\begin{equation}
\overline{\langle \s_0\rangle\langle\s_{\ell}\rangle\ Z^n_{\ell,o}} = \frac{1}{n}\left[\sum_{\lambda\in\dq}\lambda^\ell(n)\ b^2_{\lambda,0}(n) - \sum_{\lambda\in\dw} \lambda^\ell(n)\ a^2_{\lambda,1}(n),\right] 
\label{cav-discn}
\end{equation}
where the coefficient $b_{\lambda,0}(n) \equiv \frac{\partial\, a_{\lambda,0}(n)}{\partial H_{0/\ell}}$ reads 
\begin{equation}
\begin{aligned}
b_{\lambda,0}(n) = \mathbb{E}_{\hh}\int \dd u\ \left[\frac{\cosh(\beta(u+\hh))}{\cosh(\beta u)}\right]^n \tanh(\beta (u+\hh)) \,\sqrt{n}\,g_0^\lambda(u;n). 
\end{aligned}
\label{b0n}
\end{equation}
We included a factor $\sqrt{n}$ in the definition of $b_{\lambda,0}(n)$ to facilitate the extrapolation of Eq. \eqref{cav-discn} to small $n$. In fact for all but the first eigenfunctions of $\dq$ the normalization condition imposes the scaling $g_0^\lambda(u;n)\sim \frac{1}{\sqrt{n}}[g_0^\lambda(u) + n\, \tilde{g}_0^\lambda(u)]$. 

We could derive Eq. \eqref{cav-discn} also in the RTM formalism for integer values of $n$ and then perform an analytic continuation to arbitrary real $n$. In the limit $n\downarrow 0$ it is easy to see that the contribution to $\overline{\langle \s_0\rangle\langle\s_{\ell}\rangle}$ from the first and the second sums of Eq. \eqref{cav-discn} are  given in Eqs. \eqref{disc-contr-0} and \eqref{disc-contr-1} of Section \ref{subsec-corr} respectively.

An alternative probabilistic derivation of the formula $\eqref{c-disc2}$ for $\overline{\langle \s_0\rangle\langle\s_{\ell}\rangle}$, which does not require the knowledge of the moments of the partition function and of $\overline{\langle \s_0\s_{\ell}\rangle_c \ Z^n_{\ell,o}}$, goes through the definition of 
\begin{equation}
 R^{(\ell)}(u) \equiv \overline{\delta(u-u_\ell)\langle\s_0\rangle^{(\ell)}}.
\end{equation}
We used the symbol $\langle\s_0\rangle^{(\ell)}$ to denote the magnetization of the first spin at the  $\ell$-th iteration of the construction of the asymmetric chain described in Section \ref{sub-cav-open}. It can be easily shown that the knowledge of $R^{(\ell)}(u)$ allows the computation of $\overline{\langle \s_0\rangle\langle\s_{\ell}\rangle}$.
Since $\langle\s_0\rangle^{(\ell)}$ is given by the derivative of the free energy of the chain at the step $\ell$ with respect to a field on the the first spin, considering the free energy difference after an iteration it is easy to arrive to the relation
\begin{equation}
\langle\s_0\rangle^{(\ell+1)}=\langle\s_0\rangle^{(\ell)}+
\left[\tanh(\beta(u_\ell+h))-\tanh(\beta u_\ell)\right]\frac{\partial u_\ell}{\partial h_0}.
\label{rec-s0}
\end{equation}
Therefore the recursion rule for $R^{(\ell)}(u)$ is given by
\begin{equation}
\begin{aligned}
R^{(\ell+1)}(u)\ =&\ \mathbb{E}_{J,h}\int \dd v\ R^{(\ell)}(v)\ \delta\left(u-\hat{u}(J,h+v)\right)\\
&+\mathbb{E}_{J,h}\int \dd v\ G_1^{(\ell)}(v)\,\delta\left(u-\hat{u}(J,h+v)\right)[\tanh(\beta(v+h))-\tanh(\beta v)],
\end{aligned}
\label{recurs-r}
\end{equation}
where $G_1^{(\ell)}(v)$ was defined in Eq. \eqref{Glq} in last paragraph.
Last equation can be solved decomposing $R^{(\ell)}(u)$  along the eigenfunctions of $\dq$ at $n=0$, 
and using Eq. \eqref{Glq} for $G_1^{(\ell)}(v)$. The computation is lengthy and not trivial, since it involves expressing $\tilde{g}^\lambda_0$ and  $\tilde{g}^\lambda_1$ (defined in Section \ref{subsec-corr}) respectively in terms of the basis of $\dq$ and $\dw$ at $n=0$. In the end though one arrives at the expression \eqref{c-disc2} for the disconnected correlation function.

\section{Conclusions}
In the present paper we presented a thorough analysis of the spectral properties of the RTM. We have developed a formalism that is suitable to  compute  many different types of connected and disconnected correlation functions and can be applied both to one-dimensional systems and to locally tree-like graphs. The expressions we found are exact for any value $\ell$ of the spin distance and can be approximated numerically considering only the top eigenvalues of certain integral operators. Also the formalism can be trivially adapted to perform the same computations in diluted p-spin models. 

We also managed to obtain exact formulas for the moments of the partition function and of the average free energies of open and closed chains of finite length. It has been recently found that short chains have an important role in the finite size corrections to disordered models on diluted graphs \cite{Ferrari2013} and in perturbative expansions around the Bethe approximation on Euclidean systems\cite{Parisi2012}. Therefore the analytical tools we have developed also apply to  these contexts.

Most of the results have also been derived using  rigorous probabilistic arguments. This approach has the merits of avoiding the complication of the decomposition of the replicated space $\ztn$ and  of being more physically intuitive than the replica one. The advantage of the replica method instead is that once the spectral representation of the RTM  is obtained all the observables can be computed just with opportune contraction. In the cavity analysis an ad-hoc iterative function or a computation strategy has to be devised for each observable.

Noticeably we did not manage to derive Eq. \eqref{f-clos} for the free energy of closed chains using a cavity argument. This is the only point withstanding the proof of the complete equivalence between the two methods.

A limitation of both the RTM formalism and of its cavity counterpart, is the fact that it is applicable to the analysis of disordered Ising models only in their replica symmetric phase. This includes all isolated one-dimensional systems but not diluted models in the spin glass phase. Therefore an investigation of the spectral properties of the 1RSB replicated transfer matrix, extending Wigner's decomposition\cite{Wigner1959} to the 1RSB symmetry group, is desirable.
Another direction for the extension of our results, which should not require too much analytical effort\cite{Weigt1997}, is toward the investigation of Potts models.

C.L. acknowledges the European
Research Council (ERC)  for financial support through grant agreement No. 247328. 

\bibliography{bibliography}

\begin{thebibliography}{33}
\expandafter\ifx\csname natexlab\endcsname\relax\def\natexlab#1{#1}\fi
\expandafter\ifx\csname bibnamefont\endcsname\relax
  \def\bibnamefont#1{#1}\fi
\expandafter\ifx\csname bibfnamefont\endcsname\relax
  \def\bibfnamefont#1{#1}\fi
\expandafter\ifx\csname citenamefont\endcsname\relax
  \def\citenamefont#1{#1}\fi
\expandafter\ifx\csname url\endcsname\relax
  \def\url#1{\texttt{#1}}\fi
\expandafter\ifx\csname urlprefix\endcsname\relax\def\urlprefix{URL }\fi
\providecommand{\bibinfo}[2]{#2}
\providecommand{\eprint}[2][]{\url{#2}}

\bibitem[{\citenamefont{Ruján}(1978)}]{Rujan78}
\bibinfo{author}{\bibfnamefont{P.}~\bibnamefont{Ruján}},
  \bibinfo{journal}{Physica A: Statistical Mechanics and its Applications}
  \textbf{\bibinfo{volume}{91}}, \bibinfo{pages}{549 } (\bibinfo{year}{1978}),
  ISSN \bibinfo{issn}{0378-4371},
  \urlprefix\url{http://www.sciencedirect.com/science/article/pii/0378437178901978}.

\bibitem[{\citenamefont{Derrida et~al.}(1978)\citenamefont{Derrida, Vannimenus,
  and Pomeau}}]{Derrida78}
\bibinfo{author}{\bibfnamefont{B.}~\bibnamefont{Derrida}},
  \bibinfo{author}{\bibfnamefont{J.}~\bibnamefont{Vannimenus}},
  \bibnamefont{and} \bibinfo{author}{\bibfnamefont{Y.}~\bibnamefont{Pomeau}},
  \bibinfo{journal}{Journal of Physics C: Solid State Physics}
  \textbf{\bibinfo{volume}{11}}, \bibinfo{pages}{4749} (\bibinfo{year}{1978}),
  \urlprefix\url{http://stacks.iop.org/0022-3719/11/i=23/a=019}.

\bibitem[{\citenamefont{Derrida and Hilhorst}(1983)}]{Derrida83}
\bibinfo{author}{\bibfnamefont{B.}~\bibnamefont{Derrida}} \bibnamefont{and}
  \bibinfo{author}{\bibfnamefont{H.~J.} \bibnamefont{Hilhorst}},
  \bibinfo{journal}{Journal of Physics A: Mathematical and General}
  \textbf{\bibinfo{volume}{16}}, \bibinfo{pages}{2641} (\bibinfo{year}{1983}),
  \urlprefix\url{http://stacks.iop.org/0305-4470/16/i=12/a=013}.

\bibitem[{\citenamefont{Grinstein and Mukamel}(1983)}]{Grinstein83}
\bibinfo{author}{\bibfnamefont{G.}~\bibnamefont{Grinstein}} \bibnamefont{and}
  \bibinfo{author}{\bibfnamefont{D.}~\bibnamefont{Mukamel}},
  \bibinfo{journal}{Phys. Rev. B} \textbf{\bibinfo{volume}{27}},
  \bibinfo{pages}{4503} (\bibinfo{year}{1983}),
  \urlprefix\url{http://link.aps.org/doi/10.1103/PhysRevB.27.4503}.

\bibitem[{\citenamefont{Nieuwenhuizen and Luck}(1986)}]{Nieuwenhuizen86}
\bibinfo{author}{\bibfnamefont{T.~M.} \bibnamefont{Nieuwenhuizen}}
  \bibnamefont{and} \bibinfo{author}{\bibfnamefont{J.~M.} \bibnamefont{Luck}},
  \bibinfo{journal}{Journal of Physics A: Mathematical and General}
  \textbf{\bibinfo{volume}{19}}, \bibinfo{pages}{1207} (\bibinfo{year}{1986}),
  \urlprefix\url{http://stacks.iop.org/0305-4470/19/i=7/a=022}.

\bibitem[{\citenamefont{Luck and Nieuwenhuizen}(1989)}]{Luck89}
\bibinfo{author}{\bibfnamefont{J.~M.} \bibnamefont{Luck}} \bibnamefont{and}
  \bibinfo{author}{\bibfnamefont{T.~M.} \bibnamefont{Nieuwenhuizen}},
  \bibinfo{journal}{Journal of Physics A: Mathematical and General}
  \textbf{\bibinfo{volume}{22}}, \bibinfo{pages}{2151} (\bibinfo{year}{1989}),
  \urlprefix\url{http://stacks.iop.org/0305-4470/22/i=12/a=017}.

\bibitem[{\citenamefont{Luck et~al.}(1991)\citenamefont{Luck, Funke, and
  Nieuwenhuizen}}]{Luck91}
\bibinfo{author}{\bibfnamefont{J.~M.} \bibnamefont{Luck}},
  \bibinfo{author}{\bibfnamefont{M.}~\bibnamefont{Funke}}, \bibnamefont{and}
  \bibinfo{author}{\bibfnamefont{T.~M.} \bibnamefont{Nieuwenhuizen}},
  \bibinfo{journal}{Journal of Physics A: Mathematical and General}
  \textbf{\bibinfo{volume}{24}}, \bibinfo{pages}{4155} (\bibinfo{year}{1991}),
  \urlprefix\url{http://stacks.iop.org/0305-4470/24/i=17/a=030}.

\bibitem[{\citenamefont{Derrida et~al.}(1986)\citenamefont{Derrida,
  Mendès~France, and Peyrière}}]{Derrida86}
\bibinfo{author}{\bibfnamefont{B.}~\bibnamefont{Derrida}},
  \bibinfo{author}{\bibfnamefont{M.}~\bibnamefont{Mendès~France}},
  \bibnamefont{and}
  \bibinfo{author}{\bibfnamefont{J.}~\bibnamefont{Peyrière}},
  \bibinfo{journal}{Journal of Statistical Physics}
  \textbf{\bibinfo{volume}{45}}, \bibinfo{pages}{439} (\bibinfo{year}{1986}),
  ISSN \bibinfo{issn}{0022-4715},
  \urlprefix\url{http://dx.doi.org/10.1007/BF01021080}.

\bibitem[{\citenamefont{Forgacs et~al.}(1984)\citenamefont{Forgacs, Mukamel,
  and Pelcovits}}]{Forgacs84}
\bibinfo{author}{\bibfnamefont{G.}~\bibnamefont{Forgacs}},
  \bibinfo{author}{\bibfnamefont{D.}~\bibnamefont{Mukamel}}, \bibnamefont{and}
  \bibinfo{author}{\bibfnamefont{R.~A.} \bibnamefont{Pelcovits}},
  \bibinfo{journal}{Phys. Rev. B} \textbf{\bibinfo{volume}{30}},
  \bibinfo{pages}{205} (\bibinfo{year}{1984}),
  \urlprefix\url{http://link.aps.org/doi/10.1103/PhysRevB.30.205}.

\bibitem[{\citenamefont{Skantzos and Coolen}(2000)}]{Skantzos00}
\bibinfo{author}{\bibfnamefont{N.~S.} \bibnamefont{Skantzos}} \bibnamefont{and}
  \bibinfo{author}{\bibfnamefont{A.~C.~C.} \bibnamefont{Coolen}},
  \bibinfo{journal}{Journal of Physics A: Mathematical and General}
  \textbf{\bibinfo{volume}{33}}, \bibinfo{pages}{1841} (\bibinfo{year}{2000}),
  \urlprefix\url{http://stacks.iop.org/0305-4470/33/i=9/a=309}.

\bibitem[{\citenamefont{Fisher et~al.}(2001)\citenamefont{Fisher, Le~Doussal,
  and Monthus}}]{Fisher01}
\bibinfo{author}{\bibfnamefont{D.~S.} \bibnamefont{Fisher}},
  \bibinfo{author}{\bibfnamefont{P.}~\bibnamefont{Le~Doussal}},
  \bibnamefont{and} \bibinfo{author}{\bibfnamefont{C.}~\bibnamefont{Monthus}},
  \bibinfo{journal}{Phys. Rev. E} \textbf{\bibinfo{volume}{64}},
  \bibinfo{pages}{066107} (\bibinfo{year}{2001}),
  \urlprefix\url{http://link.aps.org/doi/10.1103/PhysRevE.64.066107}.

\bibitem[{\citenamefont{Coolen and Takeda}(2012)}]{Coolen12}
\bibinfo{author}{\bibfnamefont{A.~C.} \bibnamefont{Coolen}} \bibnamefont{and}
  \bibinfo{author}{\bibfnamefont{K.}~\bibnamefont{Takeda}},
  \bibinfo{journal}{Philosophical Magazine} \textbf{\bibinfo{volume}{92}},
  \bibinfo{pages}{64} (\bibinfo{year}{2012}),
  \eprint{http://dx.doi.org/10.1080/14786435.2011.606238},
  \urlprefix\url{http://dx.doi.org/10.1080/14786435.2011.606238}.

\bibitem[{\citenamefont{Weigt and Monasson}(1996)}]{Monasson96}
\bibinfo{author}{\bibfnamefont{M.}~\bibnamefont{Weigt}} \bibnamefont{and}
  \bibinfo{author}{\bibfnamefont{R.}~\bibnamefont{Monasson}},
  \bibinfo{journal}{EPL (Europhysics Letters)} \textbf{\bibinfo{volume}{36}},
  \bibinfo{pages}{4} (\bibinfo{year}{1996}), \eprint{9608149},
  \urlprefix\url{http://stacks.iop.org/0295-5075/36/i=3/a=209
  http://arxiv.org/abs/cond-mat/9608149}.

\bibitem[{\citenamefont{Nikoletopoulos and Coolen}(2004)}]{Nikoletopoulos2004}
\bibinfo{author}{\bibfnamefont{T.}~\bibnamefont{Nikoletopoulos}}
  \bibnamefont{and} \bibinfo{author}{\bibfnamefont{A.~C.~C.}
  \bibnamefont{Coolen}}, \bibinfo{journal}{Journal of Physics A: Mathematical
  and General} \textbf{\bibinfo{volume}{37}}, \bibinfo{pages}{8433}
  (\bibinfo{year}{2004}), ISSN \bibinfo{issn}{0305-4470}, \eprint{0405269v2},
  \urlprefix\url{http://stacks.iop.org/0305-4470/37/i=35/a=003?key=crossref.1e760bc99a921a1e3113518a33fbe36b}.

\bibitem[{\citenamefont{Parisi et~al.}(1987)\citenamefont{Parisi, M\'{e}zard,
  and Virasoro}}]{Parisi1987}
\bibinfo{author}{\bibfnamefont{G.}~\bibnamefont{Parisi}},
  \bibinfo{author}{\bibfnamefont{M.}~\bibnamefont{M\'{e}zard}},
  \bibnamefont{and} \bibinfo{author}{\bibfnamefont{M.~A.}
  \bibnamefont{Virasoro}}, \emph{\bibinfo{title}{{Spin glass theory and
  beyond}}} (\bibinfo{publisher}{World Scientific Singapore},
  \bibinfo{year}{1987}).

\bibitem[{\citenamefont{Montanari and M\'{e}zard}(2009)}]{Montanari2009}
\bibinfo{author}{\bibfnamefont{A.}~\bibnamefont{Montanari}} \bibnamefont{and}
  \bibinfo{author}{\bibfnamefont{M.}~\bibnamefont{M\'{e}zard}},
  \emph{\bibinfo{title}{{Information, Physics and Computation}}}
  (\bibinfo{publisher}{Oxford Univ. Press}, \bibinfo{year}{2009}).

\bibitem[{\citenamefont{Young}(1998)}]{Young98}
\bibinfo{author}{\bibfnamefont{A.}~\bibnamefont{Young}},
  \emph{\bibinfo{title}{Spin Glasses and Random Fields}}, Directions in
  condensed matter physics (\bibinfo{publisher}{World Scientific},
  \bibinfo{year}{1998}), ISBN \bibinfo{isbn}{9789810232405},
  \urlprefix\url{http://books.google.it/books?id=9hsCr-i6b5wC}.

\bibitem[{\citenamefont{Ferrari et~al.}(2013)\citenamefont{Ferrari, Lucibello,
  Morone, Parisi, Ricci-Tersenghi, and Rizzo}}]{Ferrari2013}
\bibinfo{author}{\bibfnamefont{U.}~\bibnamefont{Ferrari}},
  \bibinfo{author}{\bibfnamefont{C.}~\bibnamefont{Lucibello}},
  \bibinfo{author}{\bibfnamefont{F.}~\bibnamefont{Morone}},
  \bibinfo{author}{\bibfnamefont{G.}~\bibnamefont{Parisi}},
  \bibinfo{author}{\bibfnamefont{F.}~\bibnamefont{Ricci-Tersenghi}},
  \bibnamefont{and} \bibinfo{author}{\bibfnamefont{T.}~\bibnamefont{Rizzo}},
  \bibinfo{journal}{Physical Review B} \textbf{\bibinfo{volume}{88}}
  (\bibinfo{year}{2013}), ISSN \bibinfo{issn}{1098-0121},
  \urlprefix\url{http://link.aps.org/doi/10.1103/PhysRevB.88.184201}.

\bibitem[{\citenamefont{Gyorgyi and Rujan}(1984)}]{Gyorgyi84}
\bibinfo{author}{\bibfnamefont{G.}~\bibnamefont{Gyorgyi}} \bibnamefont{and}
  \bibinfo{author}{\bibfnamefont{P.}~\bibnamefont{Rujan}},
  \bibinfo{journal}{Journal of Physics C: Solid State Physics}
  \textbf{\bibinfo{volume}{17}}, \bibinfo{pages}{4207} (\bibinfo{year}{1984}),
  \urlprefix\url{http://stacks.iop.org/0022-3719/17/i=24/a=004}.

\bibitem[{\citenamefont{Aeppli and Bruinsma}(1983)}]{Aeppli83}
\bibinfo{author}{\bibfnamefont{G.}~\bibnamefont{Aeppli}} \bibnamefont{and}
  \bibinfo{author}{\bibfnamefont{R.}~\bibnamefont{Bruinsma}},
  \bibinfo{journal}{Physics Letters A} \textbf{\bibinfo{volume}{97}},
  \bibinfo{pages}{117 } (\bibinfo{year}{1983}), ISSN \bibinfo{issn}{0375-9601},
  \urlprefix\url{http://www.sciencedirect.com/science/article/pii/0375960183905285}.

\bibitem[{\citenamefont{Bruinsma and Aeppli}(1983)}]{Bruinsma1983}
\bibinfo{author}{\bibfnamefont{R.}~\bibnamefont{Bruinsma}} \bibnamefont{and}
  \bibinfo{author}{\bibfnamefont{G.}~\bibnamefont{Aeppli}},
  \bibinfo{journal}{Physical Review Letters} \textbf{\bibinfo{volume}{50}},
  \bibinfo{pages}{1494} (\bibinfo{year}{1983}), ISSN \bibinfo{issn}{0031-9007},
  \urlprefix\url{http://link.aps.org/doi/10.1103/PhysRevLett.50.1494}.

\bibitem[{\citenamefont{Behn and Zagrebnov}(1988)}]{Behn88}
\bibinfo{author}{\bibfnamefont{U.}~\bibnamefont{Behn}} \bibnamefont{and}
  \bibinfo{author}{\bibfnamefont{V.~A.} \bibnamefont{Zagrebnov}},
  \bibinfo{journal}{Journal of Physics A: Mathematical and General}
  \textbf{\bibinfo{volume}{21}}, \bibinfo{pages}{2151} (\bibinfo{year}{1988}),
  \urlprefix\url{http://stacks.iop.org/0305-4470/21/i=9/a=028}.

\bibitem[{\citenamefont{Igloi}(1994)}]{Igloi94}
\bibinfo{author}{\bibfnamefont{F.}~\bibnamefont{Igloi}},
  \bibinfo{journal}{Journal of Physics A: Mathematical and General}
  \textbf{\bibinfo{volume}{27}}, \bibinfo{pages}{2995} (\bibinfo{year}{1994}),
  \urlprefix\url{http://stacks.iop.org/0305-4470/27/i=9/a=015}.

\bibitem[{\citenamefont{Janzen and Engel}(2010)}]{Janzen2010}
\bibinfo{author}{\bibfnamefont{K.}~\bibnamefont{Janzen}} \bibnamefont{and}
  \bibinfo{author}{\bibfnamefont{A.}~\bibnamefont{Engel}},
  \bibinfo{journal}{Journal of Statistical Mechanics: Theory and Experiment}
  pp. \bibinfo{pages}{1--15} (\bibinfo{year}{2010}), \eprint{1008.1733v1}.

\bibitem[{\citenamefont{Janzen et~al.}(2010)\citenamefont{Janzen, Engel, and
  M\'{e}zard}}]{Janzen2010a}
\bibinfo{author}{\bibfnamefont{K.}~\bibnamefont{Janzen}},
  \bibinfo{author}{\bibfnamefont{A.}~\bibnamefont{Engel}}, \bibnamefont{and}
  \bibinfo{author}{\bibfnamefont{M.}~\bibnamefont{M\'{e}zard}},
  \bibinfo{journal}{Physical Review E} \textbf{\bibinfo{volume}{82}}
  (\bibinfo{year}{2010}), ISSN \bibinfo{issn}{1539-3755},
  \eprint{arXiv:1006.2927v1},
  \urlprefix\url{http://link.aps.org/doi/10.1103/PhysRevE.82.021127}.

\bibitem[{\citenamefont{Parisi et~al.}(2014)\citenamefont{Parisi,
  Ricci-Tersenghi, and Rizzo}}]{Parisi14}
\bibinfo{author}{\bibfnamefont{G.}~\bibnamefont{Parisi}},
  \bibinfo{author}{\bibfnamefont{F.}~\bibnamefont{Ricci-Tersenghi}},
  \bibnamefont{and} \bibinfo{author}{\bibfnamefont{T.}~\bibnamefont{Rizzo}},
  \bibinfo{journal}{Journal of Statistical Mechanics: Theory and Experiment}
  \textbf{\bibinfo{volume}{2014}}, \bibinfo{pages}{P04013}
  (\bibinfo{year}{2014}),
  \urlprefix\url{http://stacks.iop.org/1742-5468/2014/i=4/a=P04013}.

\bibitem[{\citenamefont{Parisi}(2010)}]{Parisi2012}
\bibinfo{author}{\bibfnamefont{G.}~\bibnamefont{Parisi}}, \bibinfo{journal}{PoS
  (HRMS)}  (\bibinfo{year}{2010}), \eprint{1201.5813},
  \urlprefix\url{http://arxiv.org/abs/1201.5813v1}.

\bibitem[{\citenamefont{{De Dominicis} and Giardina}(2006)}]{DeDominicis2006}
\bibinfo{author}{\bibfnamefont{C.}~\bibnamefont{{De Dominicis}}}
  \bibnamefont{and} \bibinfo{author}{\bibfnamefont{I.}~\bibnamefont{Giardina}},
  \emph{\bibinfo{title}{{Random fields and spin glasses}}}
  (\bibinfo{publisher}{Cambridge University Press}, \bibinfo{year}{2006}).

\bibitem[{\citenamefont{Wigner}(1959)}]{Wigner1959}
\bibinfo{author}{\bibfnamefont{E.~P.} \bibnamefont{Wigner}},
  \emph{\bibinfo{title}{{Group theory and its application to the quantum
  mechanics of atomic spectra}}} (\bibinfo{year}{1959}).

\bibitem[{\citenamefont{Bray and Moore}(1979)}]{Bray1979}
\bibinfo{author}{\bibfnamefont{A.~J.} \bibnamefont{Bray}} \bibnamefont{and}
  \bibinfo{author}{\bibfnamefont{M.~A.} \bibnamefont{Moore}},
  \bibinfo{journal}{Journal of Physics C: Solid State Physics}
  \textbf{\bibinfo{volume}{12}} (\bibinfo{year}{1979}).

\bibitem[{\citenamefont{de~Almeida and Thouless}(1978)}]{Thouless1978}
\bibinfo{author}{\bibfnamefont{J.~R.~L.} \bibnamefont{de~Almeida}}
  \bibnamefont{and} \bibinfo{author}{\bibfnamefont{D.~J.}
  \bibnamefont{Thouless}}, \bibinfo{journal}{Journal of Physics A: Mathematical
  and General} \textbf{\bibinfo{volume}{11}}, \bibinfo{pages}{983}
  (\bibinfo{year}{1978}), ISSN \bibinfo{issn}{0305-4470},
  \urlprefix\url{http://stacks.iop.org/0305-4470/11/i=5/a=028?key=crossref.c94e59ef7ba0b393e3ee582dcdd0c876}.

\bibitem[{\citenamefont{Morone et~al.}(2013)\citenamefont{Morone, Parisi, and
  Ricci-Tersenghi}}]{Morone13}
\bibinfo{author}{\bibfnamefont{F.}~\bibnamefont{Morone}},
  \bibinfo{author}{\bibfnamefont{G.}~\bibnamefont{Parisi}}, \bibnamefont{and}
  \bibinfo{author}{\bibfnamefont{F.}~\bibnamefont{Ricci-Tersenghi}},
  \bibinfo{journal}{ArXiv e-prints} p.~\bibinfo{pages}{19}
  (\bibinfo{year}{2013}), \eprint{1308.2037},
  \urlprefix\url{http://arxiv.org/abs/1308.2037}.

\bibitem[{\citenamefont{Weigt}(1998)}]{Weigt1997}
\bibinfo{author}{\bibfnamefont{M.}~\bibnamefont{Weigt}},
  \bibinfo{journal}{Journal of Physics A: Mathematical and General}
  \textbf{\bibinfo{volume}{31}}, \bibinfo{pages}{951} (\bibinfo{year}{1998}),
  ISSN \bibinfo{issn}{0305-4470}, \eprint{9709009},
  \urlprefix\url{http://arxiv.org/abs/cond-mat/9709009
  http://stacks.iop.org/0305-4470/31/i=3/a=008?key=crossref.f9fab53d283238a543a765f9b34c78e5}.

\end{thebibliography}

\appendix

\end{document}